\documentclass[preprint]{aastex}
\usepackage{natbib}

\def\contunits{ergs\ s$^{-1}$\,cm$^{-2}$\,\AA$^{-1}$}
\def\Fvar{\ifmmode F_{\rm var} \else $F_{\rm var}$\fi}
\def\Rmax{\ifmmode R_{\rm max} \else $R_{\rm max}$\fi}
\def\tcent{\ifmmode \tau_{\rm cent} \else $\tau_{\rm cent}$\fi}
\def\tpeak{\ifmmode \tau_{\rm peak} \else $\tau_{\rm peak}$\fi}
\def\rmax{\ifmmode r_{\rm max} \else $r_{\rm max}$\fi}
\def\MBH{\ifmmode M_{\rm BH} \else $M_{\rm BH}$\fi}
\def\sigstar{\ifmmode \sigma_* \else $\sigma_*$\fi}
\def\sigline{\ifmmode \sigma_{\rm line} \else $\sigma_{\rm line}$\fi}
\def\Msigma{\ifmmode M_{\rm BH}$--$\sigma_* 
  \else $M_{\rm BH}$--$\sigma_*$\fi}
\def\IUE{{\it IUE}}
\def\HST{{\it HST}}

\def\arcsecpoint{$''\!.$}

\def\kms{\ifmmode {\rm km\ s}^{-1} \else km s$^{-1}$\fi}
\def\Msun{\ifmmode M_{\odot} \else $M_{\odot}$\fi}
\def\Lsun{\ifmmode L_{\odot} \else $L_{\odot}$\fi}
\def\qo{\ifmmode q_{\rm o} \else $q_{\rm o}$\fi}
\def\Ho{\ifmmode H_{\rm o} \else $H_{\rm o}$\fi}
\def\Hubble{km\,s$^{-1}$\,Mpc$^{-1}$}
\def\ho{\ifmmode h_{\rm o} \else $h_{\rm o}$\fi}

\def\vFWHM{\ifmmode v_{\mbox{\tiny FWHM}} \else
            $v_{\mbox{\tiny FWHM}}$\fi}
\def\CCF{\ifmmode F_{\it CCF} \else $F_{\it CCF}$\fi}
\def\ACF{\ifmmode F_{\it ACF} \else $F_{\it ACF}$\fi}
\def\Halpha{\ifmmode {\rm H}\alpha \else H$\alpha$\fi}
\def\Hbeta{\ifmmode {\rm H}\beta \else H$\beta$\fi}
\def\Hgamma{\ifmmode {\rm H}\gamma \else H$\gamma$\fi}
\def\Hdelta{\ifmmode {\rm H}\delta \else H$\delta$\fi}
\def\Lya{\ifmmode {\rm Ly}\alpha \else Ly$\alpha$\fi}
\def\Lyb{\ifmmode {\rm Ly}\beta \else Ly$\beta$\fi}

\def\hei{He\,{\sc i}}
\def\heii{He\,{\sc ii}}

\def\ciii{\ifmmode {\rm C}\,{\sc iii} \else C\,{\sc iii}\fi}
\def\civ{\ifmmode {\rm C}\,{\sc iv} \else C\,{\sc iv}\fi}

\def\nv{N\,{\sc v}}

\def\oiii{O\,{\sc iii}}
\def\o5007{[O\,{\sc iii}]\,$\lambda5007$}

\def\feii{Fe\,{\sc ii}}

\shorttitle{AGN Black Hole Masses}
\shortauthors{Peterson et al.}

\begin{document}
\title{Central Masses and Broad-Line Region Sizes of Active Galactic
Nuclei.\ II.\ A Homogeneous Analysis of a Large Reverberation-Mapping
Database}

\author{B.M.~Peterson\altaffilmark{1},
L.~Ferrarese\altaffilmark{2},
K.M.~Gilbert\altaffilmark{1,3},
S.~Kaspi\altaffilmark{4,5},
M.A.~Malkan\altaffilmark{6},
D.~Maoz\altaffilmark{4},
D.~Merritt\altaffilmark{7},
H.~Netzer\altaffilmark{4},
C.A.~Onken\altaffilmark{1},
R.W.~Pogge\altaffilmark{1},
M.~Vestergaard\altaffilmark{8},
and A.~Wandel\altaffilmark{9}}

\altaffiltext{1}{Department of Astronomy, The Ohio State
University, 140 West 18th Avenue, Columbus, OH  43210}
\altaffiltext{2}{Department of Physics and Astronomy,
Rutgers University, 136 Frelinghuysen Road,
Piscataway, NJ 08854}
\altaffiltext{3}{Present address: University of California Observatories,
Lick Observatory,
University of California, Santa Cruz, CA 92064}
\altaffiltext{4}{Wise Observatory and 
School of Physics and Astronomy, 
Raymond and Beverly Sackler Faculty of Exact Sciences,
Tel-Aviv University, Tel-Aviv 69978, ISRAEL}
\altaffiltext{5}{Department of Physics, Technion, Haifa 32000, ISRAEL}
\altaffiltext{6}{Department of Physics and Astronomy,
University of California, Los Angeles, CA 90095}
\altaffiltext{7}{Department of Physics, Rochester Institute of Technology,
84 Lomb Memorial Drive, Rochester, NY 14623}
\altaffiltext{8}{Steward Observatory,
University of Arizona, Tucson, AZ 85721}
\altaffiltext{9}{Racah Institute of Physics,
The Hebrew University, Jerusalem 91405, ISRAEL}

\email{peterson@astronomy.ohio-state.edu; 
lff@physics.rutgers.edu;
kgilbert@astro.ucsc.edu;
shai@wise.tau.ac.il;
malkan@bonnie.astro.ucla.edu;
dani@wise.tau.ac.il;
drmsps@ad.rit.edu;
netzer@wise.tau.ac.il;
onken@astronomy.ohio-state.edu; 
pogge@astronomy.ohio-state.edu; 
mvestergaard@as.arizona.edu;
amri@vms.huij.ac.il}

\begin{abstract}
We present improved black hole masses for 35 active galactic
nuclei (AGNs) based on a complete and consistent reanalysis of 
broad emission-line reverberation-mapping data. From objects
with multiple line measurements, we find that the highest precision
measure of the virial
product $c\tau\Delta V^2/G$, where $\tau$ is the emission-line 
lag relative to continuum variations and $\Delta V$ is the 
emission-line width, is obtained by using the cross-correlation
function centroid (as opposed to the cross-correlation function peak)
for the time delay and the line dispersion
(as opposed to full width half maximum) for the line width
and by measuring the line width in the {\em variable} part of
the spectrum. Accurate line-width measurement depends
critically on avoiding contaminating features,
in particular the narrow components of the emission lines.
We find that the precision (or random component of
the error) of reverberation-based black hole mass measurements
is typically around 30\%, comparable to the
precision attained in measurement of black hole masses in
quiescent galaxies by gas or stellar dynamical methods.
Based on results presented in a companion paper by Onken et al.,
we provide a zero-point calibration for the reverberation-based
black hole mass scale by using the relationship between
black hole mass and host-galaxy bulge velocity dispersion.
The scatter around this relationship implies that the
typical systematic uncertainties in reverberation-based black hole
masses are smaller than a factor of three.
We present a preliminary version of a mass--luminosity relationship
that is much better defined than any previous attempt.
Scatter about the mass--luminosity
relationship for these AGNs appears to be real and
could be correlated with either Eddington ratio or object inclination.

\end{abstract}


\section{INTRODUCTION}

The evidence that active galactic nuclei (AGNs) are powered by 
gravitational accretion onto supermassive black holes is now 
quite convincing. Certainly there has not yet been a
definitive detection of the relativistic effects that 
would be required for unambiguous identification of a singularity,
although studies of the Fe K$\alpha$ emission line in the
X-ray spectra of AGNs currently affords some promise
(e.g., Reynolds \& Nowak 2002). Nevertheless it seems to be true 
that the centers of both active and quiescent galaxies host 
supermassive (greater than $\sim10^6\,\Msun$) objects that must 
be so compact that other alternatives are very unlikely.  

Black hole masses are measured in a number of 
ways. In quiescent galaxies, dynamical modelling of either stellar kinematics
(e.g., van der Marel 1994; van der Marel et al.\ 1998; 
Verolme et al.\ 2002; Gebhardt et al.\ 2003)
or gas motions (e.g., Harms et al.\ 1994; Ford et al.\ 1994;
Macchetto et al.\ 1997) is used to determine central masses. In the case of 
NGC~4258, a weakly active galaxy, proper motions and radial 
velocities of H$_2$O megamaser spots are used to deduce a high 
precision central mass (Miyoshi et al.\ 1995; Herrnstein et al.\ 1999). 
In Type 1 active galaxies (i.e., those with prominent broad
emission lines in their UV/optical spectra), reverberation 
mapping (Blandford \& McKee 1982; Peterson 1993) of the 
broad-line region (BLR) can be used to determine the central 
masses. Reverberation mapping is the only method that does 
not depend on high angular resolution, so it is of special interest 
as it is thus extendable in principle to both very high and very 
low luminosities and to objects at great distances. Moreover, 
reverberation studies reveal the existence of simple scaling 
relationships that can be used to anchor secondary methods of 
mass measurement, thus making it possible to provide estimates 
of the masses of large samples of quasars, including
even very distant quasars, based on  relatively simple 
spectral measurements (e.g., Vestergaard 2002, 2004; McLure \& Jarvis 2002).

While reverberation methods in principle can be used to 
determine the full geometry and kinematics of the BLR
(e.g., Horne et al.\ 2004), 
applications to date have been comparatively simple. Time 
delays between continuum and emission-line variations are used 
to deduce the size of the BLR, or more accurately, the size of the 
line-emitting region for the particular emission line in question. 
By combining the measured time delay $\tau$ with the 
emission-line width $\Delta V$, a virial mass can be obtained,
\begin{equation}
\label{eq:virial1}
M = \frac{f c\tau \Delta V^2}{G},
\end{equation}
where $c$ is the speed of light and $G$ is the gravitational 
constant. The factor $f$ is of order unity and depends on the 
structure, kinematics, and orientation of the BLR. 
We will sometimes refer below to the ``unscaled'' virial
mass $M_{p} = c\tau \Delta V^2/G$, as the virial product,
so that the virial mass is the virial product times the
scaling factor $f$. 

It became clear 
even in the first well-sampled reverberation program on NGC~5548 
(Clavel et al.\ 1991; Peterson et al.\ 1991; Dietrich et al.\ 
1993; Maoz et al.\ 1993) 
that different emission lines have different time-delayed 
responses, or lags. Lags are shorter for lines that are
characteristic of more highly ionized gases, i.e., the BLR has a 
stratified ionization structure. It was already known 
(e.g., Osterbrock \& Shuder 1982) that higher 
ionization lines (e.g., \heii\,$\lambda4686$) are broader than 
lower ionization lines  (e.g., \Hbeta\,$\lambda4861$), 
and it was natural to look 
for a virial relationship between lag and line width, $\tau \propto 
\Delta V^{-2}$, which would constitute evidence that gravity 
dominates the motions of the BLR gas and that 
the black hole mass can therefore be inferred. Early attempts to do 
this were not promising, although Krolik et al.\ (1991)
did note the trend of decreasing time lag with increasing line
width for the UV lines in NGC~5548. However, 
upon revisiting the issue, Peterson \& Wandel (1999) found that 
there is indeed a virial relationship between lag and line width in 
the case of NGC 5548 if the line width is measured in the {\em 
variable} part of the emission line and one avoids (a) lines that are 
strongly blended with other features and (b) lines with lags that are 
uncertain because of potential aliasing effects in the time-series 
analysis. Similar virial-like relationships between lag and line 
width were subsequently found in other objects (Peterson \& 
Wandel 2000; Onken \& Peterson 2002; Kollatschny 2003). 
Despite earlier claims that emission-line 
reverberation yielded masses that were too low by 
a factor of several (Ho 1999), 
it was subsequently shown (Gebhardt et al. 2000b; 
Ferrarese et al.\ 2001) that the relationship between AGN 
reverberation-based black hole masses \MBH\ and their 
host-galaxy bulge velocity dispersions \sigstar\ appears to be 
consistent with the black hole mass/bulge velocity dispersion 
relationship (hereafter the \Msigma\ relationship; Ferrarese \& 
Merritt 2000; Gebhardt et al.\ 2000a) that is seen in normal 
galaxies. Moreover, the relationship between black hole
mass and host-galaxy bulge luminosity also seems to be the same
for both quiescent and active galaxies (Wandel 2002;
McLure \& Dunlop 2002).

Unfortunately, there is a significant systematic uncertainty 
(Krolik 2001) in AGN reverberation masses 
embodied in the scaling factor $f$ in 
eq.\ (\ref{eq:virial1}), which remains unknown.
For lack of a better estimate,
published studies have usually used a nominal value of $f=0.75$
for $\Delta V$ taken to be the full width half maximum (FWHM)
of the emission line, as described in section 6 below.

Wandel, Peterson, \& Malkan (1999; Paper I in this series and 
hereafter referred to as WPM) published a compilation of 
black hole masses in 17 Seyfert 1 galaxies and two quasars. 
Kaspi et al.\ 
(2000) published the results of a large reverberation-mapping 
campaign that led to mass measurements for the central objects 
in 17 Palomar-Green (Schmidt \& Green 1983) quasars and combined 
their results with WPM to obtain relationships between the BLR 
size and AGN optical luminosity (the ``radius--luminosity relationship'') 
and between the central mass, which we will henceforth assume 
to be a black hole, and the optical luminosity 
(the ``mass--luminosity relationship''). 
These are both of obvious importance: 
\begin{enumerate}
\item The radius-luminosity relationship can be used to deduce 
the masses of black holes in distant quasars by combining the 
inferred BLR radius with the widths of the emission lines.
\item The mass-luminosity relationship hence obtained relates directly to 
current accretion rates and radiative efficiencies.
The mass itself provides a strong 
constraint on the black hole growth history.
\end{enumerate}

All of the important relationships mentioned here --- 
the time-lag/line-width virial relationship, the 
AGN \Msigma\ relationship, the 
radius--luminosity relationship, and the mass--luminosity 
relationship ---  show considerable scatter.
Moreover, the reverberation database is very 
inhomogeneous, and the data have not always been analyzed in a 
uniform way; this is particularly true in the case of many of the 
earliest results. We suspected that more than a decade of 
experience in developing reverberation mapping techniques and 
error analysis merited reanalysis of the earlier data and, that
in at least some cases, improved calibration for spectra
would lead to improved 
results. This suspicion was borne out in the case of NGC 3783 
(Onken \& Peterson 2002); our reanalysis of the combined UV
and optical data led to a determination 
of the virial mass of the central object that was an order of
magnitude  more precise than that quoted by WPM, which was based only on the 
original optical spectra and optical continuum and \Hbeta\ light curves.
Fundamentally, a complete reanalysis of the body of reverberation
database is warranted by relatively recent 
(a) improvements in cross-correlation error analysis, 
(b) recognition of the importance of measuring
line widths in the {\em variable} part of the spectrum, and
(c) recognition that emission-line time lags can vary over
time scales longer than the reverberation time scale due
to changes in the mean luminosity of the object (Peterson et al.\ 2002).

We thus decided to undertake a massive reanalysis of all of the 
reverberation mapping data readily available to us, for the 
express purpose of improving AGN black hole mass 
determinations. We distinguish 
in the usual way (e.g., Bevington 1969) between the
{\em accuracy} of these masses (i.e., how close they are
to the true values), which depends on how
well we can account for systematics,
and their {\em precision} (i.e., how exactly we measure
the virial products), which depends primarily on
``random errors'' associated with measuring line widths
and time lags. Thus, this investigation has two parts, with different 
goals:
\begin{enumerate}
\item In order to improve the precision of AGN black hole 
mass measurements, we are reanalyzing all of the readily 
available reverberation data to determine the best measures of 
time lag and line width for these studies. We do this by assuming 
that the virial product for all emission lines is the same at all 
times for a particular AGN. 
We then explore ways of characterizing lags and line 
widths that yield relationships most consistent with a virial 
relationship. This is the subject of this contribution.
\item In order to improve the accuracy of AGN black hole 
mass measurements, we are obtaining high-precision 
measurements of \sigstar\ for reverberation-mapped AGNs.
We demonstrate that there is general consistency between the 
\Msigma\ relationships between AGNs and quiescent galaxies, 
and then assume that these two relationships have a common 
zero point, thus determining the scale factor 
$f$. This is the subject of a companion 
paper (Onken et al.\ 2004), which we will draw on for the 
absolute calibration of the black hole mass scale.
\end{enumerate}

\section{DATA}
We have included in this analysis all objects for
which we had ready access to the spectra used
in the original investigations. This consists of data from
most of the reverberation-mapping experiments undertaken to date,
including the large samples from 
International AGN Watch\footnote{Data obtained as part of 
International AGN Watch projects are available at
http://www.astronomy.ohio-state.edu/$\sim$agnwatch/.}
projects (Alloin et al.\ 1994; Peterson 1999),
the Lovers of Active Galaxies (LAG) campaign (Robinson 1994),
the Ohio State monitoring program (Peterson et al.\ 1998a),
and the Wise Observatory/Steward Observatory monitoring program
(Kaspi et al.\ 2000). A list of objects analyzed here is
given in Table 1. Column (1) gives the common name of the object
as used in the relevant papers on the reverberation results,
which are referenced in column (2).
Epoch 2000 coordinates are given in columns (3) and (4).
The redshift $z$ of each object is in column (5), with 
nominal $V$ magnitude and $B$-band extinction,
based on 100\,\micron\ dust maps from Schlegel,
Finkbeiner, \& Davis (1998), are given in columns (6) and (7),
respectively. Column (8) gives the standard
name of the object in the V\'{e}ron-Cetty \& V\'{e}ron (2001)
catalog, and column (9) gives other common names by which the
object is often known in the AGN literature. All entries in
this table are from the NASA/IPAC Extragalactic 
Database.

For this analysis, the fundamental data that
we require are (1) continuum and emission-line light curves
to determine time lags and (2) spectra from which line widths are to
be measured. We use the published versions of the
light curves, except where noted below. For the purpose of measuring
the line widths, we use the same spectra from which
the continuum and emission-line measurements were made; in some
cases, notably the ground-based component of the International
AGN Watch monitoring programs on Fairall 9,
NGC 3783, NGC 4051, NGC 4151, NGC 5548, 3C 390.3, and NGC 7469,
we restricted consideration to the single largest homogeneous
data subsets, i.e., those data which are most similar in terms
of resolution and quality, often from a single source.
For each set of spectra, we formed a mean spectrum,
\begin{equation}
\label{eq:meanspec}
\overline{F(\lambda)} = \frac{1}{N} \sum_{i=1}^N F_i(\lambda),
\end{equation}
where $F_i(\lambda)$ is the $i$th spectrum of the
$N$ spectra that comprise the database.
We also define  a root-mean-square (rms) spectrum
\begin{equation}
\label{eq:rmsspec}
S(\lambda) = \left[ \frac{1}{N - 1} 
\sum_{i=1}^N \left( F_i(\lambda) - \overline{F(\lambda)} \right)^2 
\right]^{1/2}.
\end{equation}
We form the mean and rms spectra only from the most
homogeneous subsets of the database, taking care to ensure that
variations observed in the homogeneous subset are consistent
with those observed in the entire data set.
Strict homogeneity of the data, particularly in terms of spectral resolution
and spectrograph entrance aperture, is necessary to avoid introduction
of spurious features in the mean and rms spectra.

Line widths can be measured in either the mean or rms spectrum;
the advantage of using the rms spectrum is that constant components
of the spectrum,
or those that vary on timescales much longer than the duration
of the experiment, vanish, thus largely
obviating the problem of deblending lines. The corresponding
disadvantage, however, is that the rms spectra are generally
much noisier than the mean spectra as the amplitude of
variability is usually fairly small for these AGNs.
The most compelling reason to use the rms spectra
is that then we are measuring the parts of the emission lines
that are actually varying.
We thus have a strong prejudice towards using the rms spectra,
and will attempt to justify this choice below.

For several of the galaxies listed in Table 1, there are multiple
data sets available, sometimes from the same source and sometimes from
different sources. We analyze each individual 
set as an independent time series.
In the case of some of the brighter Seyfert 1 galaxies which
have relatively short \Hbeta\ response times, multiple-year
campaigns were broken down into individual subsets covering
single observing seasons, thus yielding multiple, independent
measurements of the line widths, lags, and virial masses.
This is desirable not only from a statistical point of view,
but because it is now clear (Peterson et al.\ 2002) that
both lags and line widths can vary from one observing season to
the next as the mean continuum luminosity slowly varies.

\section{LINE WIDTHS}

In Table 2, we identify each individual data set for which time-series
analysis was carried out. Column (1) gives the common name
of the object, and column (2) gives the reference for the original
data. Individual emission lines are identified in column (3)
and the spectral resolution of the data (see below)
appears in column (4). Column (5) gives the range in Julian dates
spanned by the spectra\footnote{It should be noted that in many
cases, isolated points at the beginning or end of the original time series
may be excluded from our analysis.}. Individual emission lines were isolated
by interpolating a linear continuum (in units of
ergs s$^{-1}$ cm$^{-2}$\,\AA$^{-1}$) between continuum windows on
either side of the line (columns 6 and 7) underneath the line,
whose limits are given in column (8); all wavelengths in Table 2
are in the observer's reference frame.

\subsection{Measures of Line Width}
Given an emission-line profile $P(\lambda)$ (i.e., flux
per unit wavelength above a continuum interpolated
underneath the line), we parameterized the line width in
two separate ways:

\paragraph{FWHM.} How this quantity
is measured depends on whether the line is single or double
peaked. In the case of a single-peaked line, we identify
the line peak $P(\lambda)_{\rm max}$. We then start at the
short-wavelength limit of the line (column 8 of Table 2)
and search for the $\lambda_1$ such that 
$P(\lambda_1) = 0.5P(\lambda)_{\rm max}$. We then repeat
the search starting from the line peak and moving to shorter
wavelengths to find $\lambda_2$ such that 
$P(\lambda_2) = 0.5P(\lambda)_{\rm max}$. The mean of
these two wavelengths is taken to be the 
wavelength at half-maximum flux
on the short-wavelength side of the profile.
An identical procedure is used to identify the
half-maximum point on the long-wavelength side of the line,
and the difference between these is taken to be the FWHM.
For a double-peaked
line, we define a short-wavelength peak 
$P(\lambda)_{\rm max, short}$ and a long-wavelength peak
$P(\lambda)_{\rm max, long}$. We then follow procedures
similar to those above: we define
$\lambda_1$ and $\lambda_2$ relative to the short-wavelength
peak only, and compute their mean. A similar calculation
is done on the long-wavelength side, this time relative
to $P(\lambda)_{\rm max, long}$, and the FWHM is taken to
be the separation between the calculated means of $\lambda_1$ and
$\lambda_2$ on either side of the line. 
This is illustrated in Fig.\ 1.

\paragraph{Line Dispersion.} The first moment of the
line profile is
\begin{equation}
\lambda_0 = \int \lambda P(\lambda) d\lambda / \int
P(\lambda) d\lambda.
\end{equation}
We use the second moment of the profile to define
the variance or mean square dispersion
\begin{equation}
\sigline^2(\lambda) = \langle \lambda^2 \rangle - \lambda_0^2
= \left( \int \lambda^2 P(\lambda) d\lambda / 
\int P(\lambda) d\lambda \right) - \lambda_0^2.
\end{equation}
The square root of this equation is
the line dispersion \sigline\ or root-mean square (rms) width of the line. 

Both measures have intrinsic strengths and weaknesses:
FWHM is trivial to measure, except in the case of multiple-peaked
lines or noisy data, and can even be accurately estimated graphically.
Compared to \sigline, it is also less sensitive to blending with
other lines and to the contribution from extended line wings.
On the other hand, \sigline\ is well defined for arbitrary
line profiles, less sensitive to the presence of even fairly
strong narrow-line components, and as we shall see
throughout this analysis, more accurate for low-contrast lines
and the relative uncertainties are much lower than for FWHM. 
On the other hand, \sigline\ is in some cases also problematic;
$\sigline \rightarrow \infty$ for a Lorentzian profile, for example
(though in practice, the wings of any reasonable line profile
become lost in the noise). Fromerth \& Melia (2000)
point out many of the advantages of \sigline\ relative to
FWHM and that show a virial relationship between lag and line width
is also found using \sigline\ rather than FWHM to
characterize the line width.

It is worth noting at this point that there is a simple relationship
between these two quantities for a given line profile. For a Gaussian, 
${\rm FWHM}/\sigline = 2( 2\ln 2)^{1/2} \approx 2.355$,
and for a rectangular profile (produced by emission-line clouds in circular
Keplerian orbits of fixed radius and random inclination),
${\rm FWHM}/\sigline = 2(3)^{1/2} \approx 3.464$.

\subsection{Resolution Correction}
Since some of the emission lines widths are actually rather narrow,
we need to correct each line-width measurement for the finite
resolution of the spectrograph with which the data were obtained. 
We will assume that the observed line widths $\Delta \lambda_{\rm obs}$
can be written in terms of the intrinsic 
line widths $\Delta \lambda_{\rm true}$ and the spectrograph
resolution $\Delta \lambda_{\rm res}$ as
\begin{equation}
\label{eq:rescor}
\Delta \lambda_{\rm obs}^2 \approx \Delta \lambda_{\rm true}^2
 + \Delta \lambda_{\rm res}^2.
\end{equation}
Application of this equation to obtain
$\Delta \lambda_{\rm true}$ requires knowing the resolution
at which the observations were made. In order to determine
this for the optical data used here, we relied on accurate, 
high-resolution measurements of the width of the
\o5007\ line in many of the AGNs discussed here by
Whittle (1992). Whittle's FWHM measurements for
AGNs in this study are given in Table 3, in the rest frame
of each galaxy. In order to determine the resolution of 
the data used in this study, we transformed the values in
Table 3 back to the observed frame and to wavelengths units
and assumed this to be $\Delta \lambda_{\rm true}$.
We then took our measurements of FWHM(\o5007) as
$\Delta \lambda_{\rm obs}$ and solved for
$\Delta \lambda_{\rm res}$, the FWHM resolution of the data.
These are the values given in column (4) of Table 2.
We recover $\Delta \lambda_{\rm true}$ for the broad lines
by application of eq.\ (\ref{eq:rescor}).

The absence of isolated narrow lines in the UV spectra of
AGNs precluded using this method for UV spectra. Instead,
we assume a spectral resolution of 6\,\AA\ for the 
{\em International Ultraviolet Explorer (IUE)} SWP camera and
a resolution of 1.9\,\AA\ for {\em Hubble Space Telescope (HST)}
FOS spectra (e.g., Korista et al.\ 1995).

\subsection{Narrow-Line Contamination}
As noted above, in principle constant line components should
not appear in rms spectra. In practice, we find that residual
narrow-line features, generally weak but sometimes quite strong,
appear in our rms spectra. These narrow-line residuals appear
when the data are less than ideally homogeneous, in particular,
when the line-spread function is not the same for each spectrum.
The Wise Observatory spectra are particularly prone to this
because of the method used to effect a photometric calibration
of the spectra; these data are obtained through a long, wide
slit that also contains a nearby field star that is used
for relative photometric calibration. An unfortunate side effect
of this highly accurate photometric calibration method is that the
line-spread function is not well-controlled as the target can
migrate small amounts within the slit in the dispersion direction;
while this has only a small
effect on the measured broad-line widths, the narrow-line
profiles are strongly affected and fairly strong 
narrow-line residuals can result in the rms spectrum.
Again as we will see below,
accurate narrow-line removal is critical for accurate line-width
measurement. Therefore, in cases where the rms spectra show
significant residual narrow components of
\Hbeta\ and/or [\oiii]\,$\lambda\lambda4959$, 5007, we have
subtracted these components from each spectrum before combining
them into mean and rms spectra. In such cases, we used 
\o5007\ as a template profile and shifted and flux-scaled this
profile to obtain a suitable model of the narrow lines, which
we then subtracted from each spectrum prior to forming the mean
and rms spectra. The 
[\oiii]\,$\lambda5007$/[\oiii]\,$\lambda4959$ flux ratio is
fixed at a value $\sim3$ by atomic physics, but the
$\Hbeta_{\rm narrow}$/[\oiii]\,$\lambda5007$ flux ratio is different
for each galaxy. For galaxies in which narrow-line removal was
undertaken for even some of the data, the adopted narrow-line
fluxes are shown in Table 4. Most of these values are
from Marziani et al.\ (2003), although in few cases we used
our own determinations.

Unfortunately, decomposition of the narrow lines from the spectra
is much more difficult in the \Halpha\ and \Hgamma\ regions of the
optical spectrum, and was therefore not attempted. Cases in which
this might present a problem are noted below. Finally, we note that
none of the UV data from \IUE\ or \HST\ show narrow-line residuals,
consistent with little or no narrow-line contribution to the
UV emission lines in quasars (cf.\ Wills et al. 1993).

\subsection{Line Width Uncertainties}
To determine FWHM and \sigline\ and their associated uncertainties,
we employ a bootstrap method similar to that which we use for the time-series
analysis. A given data set contains $N$ spectra. For a single
bootstrap realization, we make $N$ random selections from
this group, without regard to whether or not a particular 
spectrum has been previously selected. From these $N$ spectra
we form a mean spectrum and an rms spectrum
(eqs.\ \ref{eq:meanspec} and \ref{eq:rmsspec}), 
and fits to the underlying continuum and the line
measurements are performed on these spectra.
Multiple bootstrap realizations allow us to build up a distribution of
line-width measurements from the random sets of $N$ spectra. From these
we can compute a mean value and standard deviation for the width of
each emission line, and these are the values that we will use in
this analysis.

This method of determining the line widths and associated errors
is different than what we have done previously. Line
widths and uncertainties presented by WPM were determined less
rigorously, by comparing the measurements of FWHM
obtained by using the ``highest plausible'' and ``lowest plausible''
underlying continua. Figure 2 shows a comparison between FWHM
values and associated uncertainties from this study and
those reported by WPM; note that for this particular comparison
only, we did {\em not} adjust our measurements for spectral resolution
in order to effect a more meaningful comparison with WPM.
In general, the measurements and errors are both in good agreement.
The uncertainties we find here are on average about 12\% lower
than those quoted by WPM.

The results on two quasars in Table 1,
PG~1351+640 and PG~1704+608, were deemed to be too poor to 
retain in this analysis; the emission-line variability was
simply too weak to produce reliably measurable emission
lines in the rms spectra. We note in particular that the
problem with PG~1704+608 has already been discussed in
the literature (Boroson 2003); the \Hbeta\ line in the rms spectrum
is not in fact variable broad-line emission, but 
merely residual narrow-line emission.

\section{TIME-SERIES ANALYSIS}
The methodology we employ for measurement of time lags 
and their associated errors is
the interpolation cross-correlation method, essentially
as described by White \& Peterson (1994) and by Peterson
et al.\ (1998b, hereafter P98b), but with some modifications that
are described in the Appendix. A
complete tutorial on our cross-correlation 
methodology is provided by Peterson (2001). 
The estimates of the uncertainties are based on a 
model-independent Monte Carlo method in which a single
realization yields a cross-correlation function (CCF) whose
centroid \tcent, peak value \rmax, and peak location
\tpeak\ are measured.
As discussed in the Appendix, we compute \tcent\ using
only the points at values $r\geq 0.8\,\rmax$, where \rmax\
is the peak value of the CCF.
A large number of independent realizations is used to build
up a ``cross-correlation centroid distribution (CCCD)''
and a ``cross-correlation peak distribution (CCPD)''
(cf.\ Maoz \& Netzer 1989). We take \tcent\ and \tpeak\
to be the means of these distributions. The CCCDs and CCPDs
are generally non-Gaussian, so we define upper and lower
uncertainties separately such that 15.87\% of the realizations
yield values larger than the mean plus the upper error and
that 15.87\% of the realizations
yield values smaller than the mean minus the lower error
(i.e., the errors are $\pm1\sigma$ errors if the distribution
is Gaussian).

We carried out a cross-correlation analysis for each dataset, as
summarized in Table 5. The object is listed in column (1) and
the emission line and relevant Julian Date range are listed in
columns (2) and (3), respectively (cf.\ Table 2).
Column (4) gives the peak amplitude of the 
CCF \rmax, and columns (5) and (6)
give the noise-corrected fractional variation \Fvar\ 
(cf.\ Rodr\'{\i}guez-Pascual et al.\ 1997) of the continuum and line,
respectively, during the range of dates given in column (3).
Columns (7) and (8) give the CCF centroid \tcent\ and
CCF peak \tpeak, respectively, both in the observed frame.
Uncertainties in these quantities were estimated by employing
the model-independent 
FR/RSS method of P98b, with selected modifications suggested
by Welsh (1999), as described in the Appendix. 
The uncertainty associated with \rmax\ (column 4)
is the rms variation in this quantity for the multiple Monte Carlo
realizations. Note that entries preceded by colons (:) are
deemed to be unreliable (see the appropriate notes 
on the individual objects below), i.e., these are cases where
there may be systematic uncertainties larger than indicated
by the quoted uncertainties.

\section{TESTS OF VIRIAL RELATIONSHIPS}

Our next goal is to determine empirically which measures of
time delay (\tcent\ or \tpeak) and line width
(FWHM or \sigline) provide the most robust estimates
of the black hole masses. Specifically, we consider
which combination of these measures gives us the most
consistent or minimum variance virial product, $c \tau \Delta V^2/G$,
where $\tau$ is the time delay and $\Delta V$ is the line width.
This test can be performed on four of the objects in our sample
for which long-duration multiwavelength spectroscopy allow
measurements of a number of different variable emission lines.
By far the best and most extensive data are those on NGC~5548,
and we give these great weight in our analysis. On the other
hand, we do not give much weight to the results for 3C~390.3
on account of relatively large uncertainties in both time lags and
line widths.

Table 6 gives our measurements for the time-lag and line-width parameters.
Columns (1) and (2) identify the object and data set, in the same
order as in Tables 2 and 5. Columns (3) and (4) contain \tcent\
and \tpeak, respectively; these are the values in Table 5, now
corrected for time dilation\footnote{WPM did not 
apply a time-dilation correction
for their objectss as the redshifts are all low. Kaspi et al.\ (2000)
made this correction for their own higher-redshift objects
and the objects of WPM.} by dividing by 
$1 + z$. Line-width measurements were transformed
to the rest frame of the object and converted to line-of-sight velocities;
the values for \sigline\ and FWHM are given in columns (5) and (6),
respectively. Again, a colon preceeding an entry indicates that
we do not regard the entry to be reliable.

At this point, we make the assumption that the most robust 
measures of the time delay and line width are those that most
closely yield the virial relationship
$\Delta V \propto \tau^{-1/2}$. The justification for this
assumption is simply that a virial relationship between
time delay and line width has already been established for
several objects (Peterson \& Wandel 1999, 2000; Onken \& Peterson 2002;
Kollatschny 2003). We proceed by examining 
the four cases where multiple measurements of the virial 
product are available, giving the most weight to the results
on NGC~5548.

\subsection{Virial Relationships in Individual Objects}

\paragraph{NGC 5548.} Figure 3 shows four plots of the
virial relationship for all the lines in NGC~5548, as measured
in the rms spectra. The optical data from the International
AGN Watch (Peterson et al.\ 2002 and references therein)
are divided into subsets based on single observing seasons
separated by the several-week gap when NGC~5548 is too close
to the Sun to observe. Some experimentation revealed that
division into shorter subsets led to much larger errors
in lag measurements.
The four panels in Fig.\ 3 show the possible permutations
of the virial product using \tcent\ and \tpeak\ for the time-delay
parameter and \sigline\ and FWHM as the line-width parameter.
Fits to these data and the other virial relationships described
below are summarized in Table 7, in each case for a best-fit
slope and for a force-fit to a slope of $b=-1/2$. Fits were 
obtained using the orthogonal regression program
GaussFit\footnote{GaussFit is publicly available at
ftp://clyde.as.utexas.edu/pub/gaussfit/.} (Jefferys,
Fitzpatrick, \& McArthur 1988),
which accounts for errors in both parameters.
These data show clearly that (a) the virial relationship 
$\Delta V \propto \tau^{-1/2}$ is robust, i.e., precisely
how the time delay and line width are measured is not critical,
and (b) the least scatter in this relationship is obtained
by using \tcent\ and \sigline\ to parameterize the relationship.
This is confirmed by computing the virial product 
$c \tau \Delta V^2/G$ for each measure;
the combination of \tcent\ and \sigline\ has a mean
precision (standard deviation divided by the mean) of about 0.032, which is
lower than for the other pairs of measurements.

It is important to point out that the fractional errors in 
FWHM are rather larger than those in \sigline. A simple consequence of
this is that the $\chi^2$ statistic can be misleading, as
it is larger for \sigline\ than for FWHM.
In any case, we hasten
to point out that the scatter in these relationships is
sufficiently large that it is clear that simple virial motion
is an incomplete description of gas motions in the BLR.
Figure 4 shows the virial relationship obtained by 
using the line widths measured from the mean spectra, uncorrected
for narrow-line contamination. The deleterious effect of
the narrow-line contribution on the line width measurements
is most strongly apparent for \Hbeta, as expected.
In Fig.\ 5, we show that correcting the spectra for
narrow-line \Hbeta\ improves the result, but only somewhat.
It seems clear that the rms spectrum should be used for
these measurements.

\paragraph{NGC 3783 and NGC 7469.} Figures 6 and 7 show
the virial relationship for NGC~3783 and NGC~7469 respectively,
two of the other well-studied AGNs for which multiple measurements
of the emission-line lags are available, though in both
cases there are far fewer data than for NGC~5548.
The measurements for NGC~3783 are in excellent agreement with viral 
relationship. The results for NGC~7469 are in
poorer agreement, but the best-fit slope is within
$2\sigma$ of the virial prediction.
Again our conclusions do not hinge critically on
which time-lag and line-width measures we use.

\paragraph{3C~390.3. } Figure 8 shows the case of 3C~390.3,
which seems to afford some difficulties. The left-hand
column shows a plot of line-width measures, \sigline\
in the top panel and FWHM in the bottom panel, versus
\tcent. The apparent lack of consistency with 
a virial relationship arises because:
\begin{enumerate}
\item All the lines in this object are very broad and each of
the measured lines are contaminated by blending:
\Lya\ is blended with \nv\,$\lambda1240$,
\civ\,$\lambda1549$ is blended with \heii\,$\lambda1640$,
and  \heii\,$\lambda4686$ and \Hbeta\ are blended.
\item The emission-line lags have relatively large uncertainties
and span a comparatively limited range (the largest and
smallest lags differ by a factor of only $\sim1.5$, compared
to a factor of 7--14 for the other three objects discussed
above).
\end{enumerate}
In an attempt to circumvent the problem of line blending,
we will assume that each line is intrinsically
symmetric about its nominal wavelength.
Its true width therefore can be better estimated by measuring only
the {\em unblended} half of the line and reflecting it
about the line center. The virial relationship using
these line widths is replotted in the right-hand column of
Fig.\ 8. This gives somewhat improved consistency with the virial
relationship, especially for FWHM. However, the relatively low range in lags 
and line widths and large errors in lag measurements hardly
make this object a convincing case for a virial relationship.
Given the large systematic uncertainties in the line widths
on account of line blending
and the relatively poor precision of time-delay measurements
we do not believe that these results are
{\em inconsistent} with a virial relationship. 

\subsection{General Results}

We conclude from this analysis that the most consistent
virial product is obtained by using \tcent\ and \sigline\
as the time-lag and line-width measures. The discussion
in the Appendix further assures us that \tcent\
is a good choice for the lag measurement. The virial product
computed from \tcent\ and \sigline\ is thus given for each
data set in column (7) of Table 6.

We also conclude that for the purpose of determining black hole
masses, the rms spectrum provides the most reliable line width
measurement. However, it is also clear (e.g., Fig. 5) that
the mean spectrum (or perhaps even a {\em single} spectrum),
with its much higher signal-to-noise ratio than the rms spectrum,
can be used with little penalty in accuracy, as long as one
can adequately account for contamination by other features,
notably the narrow-line contribution and blending with
adjacent features. However, the strength of narrow-line
contributions to broad-line spectra are often known rather poorly
if at all and blending by various features, notably \feii\
contamination of \Hbeta, is problematic. Further discussion
is beyond the scope of the current paper, but will be pursued
elsewhere.

For completeness, we show in Figs.\ 9 and 10
the distribution of the ratios ${\rm FWHM}/\sigline$
and $\tpeak/\tcent$, respectively, for all lines used in this analysis
(i.e., highly uncertain values excluded). The
mean and standard deviations of these distributions are
${\rm FWHM}/\sigline = 2.03 \pm 0.59$ and
$\tpeak/\tcent = 0.95 \pm 0.20$. The low mean value of
${\rm FWHM}/\sigline$ relative to that for a Gaussian
(${\rm FWHM}/\sigline = 2.355$) means that, on average,
these lines have weaker cores and stronger wings relative to Gaussians.

\section{BLACK HOLE MASSES}

For many of the objects in this study, we have
multiple measurements of the virial products.
In Table 8, we list for each object in this study
the weighted mean virial product
$<c\tcent\sigline^2/G>$, based on the entries in
column (7) of Table 6, excluding the more
uncertain values (those preceeded by a colon).

The masses of the central objects are given by
\begin{equation}
\label{eq:virial2}
M_{\rm BH} = \frac{f c\tcent \sigline^2}{G},
\end{equation}
where as noted earlier $f$ depends on the
structure, kinematics, and aspect of the BLR.
The scaling factor $f$ can be determined in a number of
ways, the easiest being to assume that 
AGNs and quiescent galaxies follow the same  \Msigma\
relationship; one can then use quiescent galaxy results 
to normalize the AGN \Msigma\ relationship and provide
an absolute mass scale for AGN black holes. We carry
out this exercise in a companion paper (Onken et al.\ 2004),
in which we find $\langle f \rangle = 5.5$.
Our final black hole masses, based on eq.\ (\ref{eq:virial2})
with an adopted mean value $\langle f \rangle = 5.5$,
are given in column (3) of Table 8.

\subsection{Uncertainties in Black Hole Masses}
As noted earlier, the first goal of this project has been
to improve the precision of the virial product measurement.
We find that the typical precision (i.e., fractional error)
of the virial product measurement is about 33\% for the 35 AGNs
for which we are able to estimate black hole masses,
or $\sim26$\%, excluding NGC~4593 and IC 4329A, for
which the reverberation results are notably poor.
The second goal is to improve the {\em statistical
accuracy} of the reverberation-based black hole mass scale
using the normalization of the AGN \Msigma\ relationship
reported in a companion paper (Onken et al.\ 2004).
The scatter around the AGN \Msigma\ relationship is
found to be about a factor 2.6 -- 2.9, depending somewhat
on the slope of the quiescent galaxy 
\Msigma\ relationship (Onken et al.\ 2004).
It is important to
keep in mind that this level of accuracy is statistical
in nature and individual black hole masses may be
less accurate.

It must be kept in mind that 
there are various systematic difficulties with
reverberation results that can in principle lead
to significant errors in individual black hole
masses (e.g., Krolik 2001). Reverberation, of course,
is not the only method of measuring black hole masses
that can fail catastrophically under certain conditions:
both stellar dynamical and gas dynamical methods
can also lead to ambiguous or even misleading results
(e.g., Verdoes Kleijn et al.\ 2000; Cappelari et al.\ 2003;
Valluri, Merritt, \& Emsellem 2004).
While we find the limited scatter in the AGN \Msigma\
relationship reassuring, additional tests remain
highly desirable.

\subsection{On Normalization of the AGN 
{\boldmath $M_{\rm BH}-\sigma_{*}$} Relationship}
As noted above, we normalize the AGN \Msigma\ relationship
to the \Msigma\ relationship for quiescent galaxies
by setting $f = 5.5$ in eq.\ (\ref{eq:virial2}).
This represents the first empirical determination
of the zero point in the AGN black hole mass
scale.

In previous work, the scale factor appeared in a 
different and more model-dependent way, 
only as an adjustment $\epsilon$ to the line-width parameter,
i.e.,
\begin{equation}
\sigma = \epsilon\, \mbox{FWHM}.
\end{equation}
and the virial mass is $M = c\langle \tau \rangle \sigma^2/G$.
For example, Netzer (1990) assumed $\epsilon = \sqrt{3}/2$,
which arises from assuming that
$\sigline = \mbox{FWHM}/2$ 
and that the velocity dispersion is isotropic,
i.e., $\sigma = \sqrt{3}\sigline$, recalling that
\sigline\ is the line-of-sight velocity dispersion. 
This further assumes that the mean time
delay $\langle\tau\rangle$ is independent of aspect or inclination,
which is true when the line emission is isotropic and unabsorbed and 
the geometry has spherical or polar symmetry.
If we make the equivalent assumptions in 
eq.\ (\ref{eq:virial2}), i.e., that the velocity
dispersion is isotropic so $\sigma = \sqrt{3}\sigline$
and that $\sigline \approx \mbox{FWHM}/2$
(which is on average quite a good assumption, given
the results of the previous section), then
$f = 3$. Our empirical calibration of the
AGN mass scale through the AGN \Msigma\ relationship
is thus about a factor of 1.8 ($=5.5/3$) higher than the mass
scale used in previous papers (e.g., WPM and Kaspi et al.\ 2000).
We emphasize again that the previous normalization was made only
in the absence of observational information or better-justified
assumptions; the value we give here is the first determination
based on observational parameters, in this case 
as embodied in the \Msigma\ relationship.

\section{COMMENTS ON SELECTED INDIVIDUAL OBJECTS}

\paragraph{PG 0026+129.} Because of strong narrow-line
residuals in the rms spectrum, we found it necessary to
remove the narrow component of \Hbeta\ and
the [\oiii]\,$\lambda\lambda 4959,$ 5007 lines from the
individual spectra. However, the narrow lines in the
vicinity of \Halpha\ are sufficiently weak that they
present no problems.

\paragraph{PG 0052+251.} The narrow-line components
were removed from the \Hbeta\ region, but the
\Halpha\ region suffers from significant narrow-line
residuals in the rms spectrum. The FWHM of \Halpha\
is untrustworthy and probably underestimated for
this reason.

\paragraph{Fairall 9.} In this case, we used the
time-binned
UV continuum light curve at 1390\,\AA\
(Rodr\'{\i}guez-Pascual et al.\ 1997) as the driving
light curve for all lines, including \Hbeta.
Unfortunately, the \civ\,$\lambda1549$ light curve
is very noisy, and the cross-correlation result
is not trustworthy. It is therefore excluded from
the mass determination.

\paragraph{Mrk 590.} 
The \Hbeta\ profile in the rms spectrum in the
first data set (JD2448090--JD2448323) seems anomalously narrow,
although the rms spectrum for this data set is significantly
noisier than for the other rms spectra
(note the lower value of \Fvar\ in Table 5). We have therefore
excluded the first data set from the mass analysis, since
the line width is questionable although the lag measurement
seems trustworthy.

\paragraph{Mrk 79.} The lag measurements from the fourth data set
(JD2449996--JD2450220) are not trustworthy on account of
some significant aliasing effects, seen clearly in
the double-peaked CCCD shown in Fig.\ 11. We have therefore
excluded this data set.

\paragraph{PG 0804+761.} This is one of a handful of
objects that are similar to the well-known Seyfert galaxy
I Zw 1. These ``I Zw 1-like'' AGNs have relatively
narrow lines and very strong optical \feii\ emission;
they are among the more extreme members of the narrow-line
Seyfert 1 (NLS1) subclass. A particular problem these objectss
present is that the [\oiii]\,$\lambda\lambda4959$, 5007
lines are heavily contaminated by \feii\ emission, and in
fact, most of the emission that makes up these features
can be attributed to \feii\,$\lambda4924$ and
\feii\,$\lambda5018$ (e.g., Peterson, Meyers, \& Capriotti 1984).
We are thus unable to use \o5007\ as a template for
removal of the narrow-line contaminants. However, in
these cases, narrow \Hbeta\ is usually too weak to strongly
affect the line-width measurements, so we do not attempt 
narrow-line removal for these objects. In this particular
quasar, there is a clear residual narrow-line \Halpha\
component in the rms spectrum, but the line-width measurement
is probably not strongly affected. On the other hand,
a strong [\oiii]\,$\lambda4363$ residual makes \Hgamma\
highly asymmetric in the rms spectrum, and thus the line-width
measurements for \Hgamma\ cannot be trusted.

\paragraph{PG 0844+349.} This is another I Zw 1-like object
(see PG 0804+761 above).
The time-lag measurements are very inconsistent from
line-to-line, probably because of inadequate
time sampling. There are narrow-line residuals in \Hgamma\
and \Halpha. The \Hbeta\ cross-correlation function is
clearly strongly affected by correlated errors and is
thus rejected. The results on this object are of rather
low quality.

\paragraph{Mrk 110.} The \Hbeta\ data used here are from
Peterson et al.\ (1998a). He\,{\sc ii}\,$\lambda4686$ appears as a prominent,
broad feature in the rms spectra of this object, as shown
in Fig.\ 12. We therefore constructed a light curve for
\heii\ from the original data. Unfortunately, the
\heii\ lags are so short that the measurements of
\tcent\ and \tpeak\ cannot be trusted, as they are significantly
shorter than the time interval between observations.
We therefore do not use the \heii\ lines in the mass determination,
although we point out that they are generally consistent
with the \Hbeta\ results. Kollatschny (2003)
has also studied the variability of this AGN, and finds
that the results for \Halpha, \Hbeta, 
\hei\,$\lambda5876$, and \heii\,$\lambda4686$ are
consistent with a single virial mass. As discussed in more
detail by Onken et al.\ (2004), the black-hole mass
measured by Kollatschny is consistent with our measurement
if we use FWHM instead of \sigline\ and Kollatschny's value
for the scaling factor $f$.

\paragraph{NGC 3227.} This AGN was the target of
two separate optical campaigns, one by the LAG
consortium in 1990 (Salamanca et al.\ 1994) and one at
CTIO in 1992 (Winge et al.\ 1995). 
The rms spectra formed from the LAG data were recently presented by
Onken et al.\ (2003). We completely reanalyzed
the CTIO data. The original reduced spectra were
rescaled in flux using the van Groningen \& Wanders (1992)
algorithm that has been used in most of the
International AGN Watch campaigns and
in the Ohio State program, and new continuum and
\Hbeta\ emission-line light curves were measured from
the rescaled spectra. While this resulted in some
improvement in the \Hbeta\ lag determination and
uncertainty, the rms spectrum was still quite noisy
because of the combination of a low amplitude of variability
and a relatively insensitive detector
(see the notes on IC 4392A, below). 

\paragraph{NGC 3516.} This is another object observed
by the LAG consortium (Wanders et al.\ 1993).
We note that the light curves for this object may
be less reliable than those of other objects as
the extended narrow-line region in this object
makes narrow line-based flux calibration vulnerable
to seeing effects, which then have to be modeled
(Wanders et al.\ 1992). We used the scaled and corrected
spectra to determine the rms spectrum, following
Onken et al.\ (2003).

\paragraph{NGC 3783.} This was the second major
multiwavelength campaign undertaken by the International
AGN Watch (Reichert et al.\ 1994; Stirpe et al.\ 1994).
This object was completely reanalyzed by Onken \& Peterson (2002);
new UV light curves based on \IUE\ NEWSIPS data
were measured, and the optical spectra were completely
recalibrated using the van Groningen \& Wanders (1992)
algorithm. The results presented here are based 
on the Onken \& Peterson reanalysis.

\paragraph{NGC 4051.} This object was also studied
recently by Shemmer et al.\ (2003), who obtain a black hole
mass consistent with the results of Peterson et al.\ (2000), 
who present the data we have used here.

\paragraph{NGC 4151.} We analyze two sets of optical
data on this object, from the Wise Observatory
campaign in 1988 (Maoz et al.\ 1991) and from
the International AGN Watch project in 1993--94
(Kaspi et al.\ 1996). Unfortunately, the rms line
profiles in the 1988 data are too poor to use due
to a combination of narrow-line residuals in the rms spectra,
uncertain narrow-line removal from the original spectra,
and a variable line-spread
function. We therefore do not use these data,
except to the extent of noting that the results
are broadly consistent with the later AGN Watch results;
though the lag measurements appear to be reliable,
the line widths in the rms spectra are not trustworthy.

\paragraph{PG 1211+143.} This AGN is another
I Zw 1-like object (see PG 0804+761 above).
Our level of confidence in the
results for this object is low: the lags are suspect
because the amplitude of variability is low, the
variations are slow, and the time sampling is not
especially good. The time lags are highly uncertain
on account of this; the CCFs are very flat-topped
and uncertain, as can be seen clearly in Fig.\ 13.
Moreover, there are some problems with
stability in the \Halpha\ region of the spectrum
that makes the line-width measurements of \Halpha\
highly uncertain. Given these difficulties, we
exclude this object from further analysis.

\paragraph{PG 1226+023.} This is the well-known quasar,
3C 273, which is another I Zw 1-like object (see PG 0804+761 above)
in which the [\oiii] lines are strongly blended with \feii\
lines (Peterson, Meyers, \& Capriotti 1984).
The narrow-line components of  \Hgamma\ and \Hbeta\
are weak in both the mean and rms spectra. However,
there are strong narrow-line residuals in the \Halpha\
region, so the \Halpha\ line-width measurements cannot
be trusted.

\paragraph{PG 1229+204.}
Inspection of the light curves shows that the variations
in this object are not well sampled. The large
differences in lags for the Balmer lines make
the results on this object rather dubious.
We have low confidence in the results for this object.

\paragraph{NGC 4593.} This is another object from the
LAG campaign in 1990 (Dietrich et al.\ 1994), where the
data have been reanalyzed by Onken et al.\ (2003).
We regard the \Hbeta\ lag as completely unreliable because
it is so much smaller than the mean sampling interval.
The \Halpha\ lag should also be regarded with some caution.

\paragraph{PG 1307+085.} The \Halpha\ region of the rms 
spectrum in this object shows residual narrow-line \Halpha, 
though this probably affects only the FWHM measurement.

\paragraph{IC 4329A.} This object and NGC 3227 were both 
observed in the CTIO monitoring 
program (Winge et al.\ 1995, 1996). These observations employed a
Reticon detector, which yielded poorer quality spectra than obtained
with the CCDs used in virtually every other ground-based campaign. We
attempted to improve the original lightcurves by 
rescaling the original spectra in flux by
using the van Groningen \& Wanders (1992) algorithm and remeasuring
the continuum and emission-line fluxes. This did improve
light curves and rms spectra, but only marginally. The light curves
are very poor and the time-lag measurements should be regarded with
caution. The FWHM measurement is very poor (note the large uncertainty
yielded by our measurement algorithm). We have little confidence in
the mass determination for this object.

\paragraph{Mrk 279.} We examined two completely 
independent sets of data, one from the Wise 
Observatory program in 1988 (Maoz et al.\ 1990) and one from an
International AGN Watch project in 1996 (Santos-Lle\'{o} et
al.\ 2001). The AGN Watch data appear to be quite good. The Wise
Observatory data, however, have a number of problems: the CCFs for both
\Halpha\ and \Hbeta\ show a correlated error signal at zero
lag\footnote{Correlated errors result from flux calibration
problems. An error in flux calibration offsets both the line and
continuum measurements based on that spectrum in the same direction,
thus introducing a spurious cross-correlation signal at zero lag.};
this problem is particularly bad at \Hbeta, and we will therefore not
use the \Hbeta\ lag measurements.  Furthermore, the \Halpha\ region of
the rms spectrum is strongly contaminated by narrow-line residuals,
and \Hbeta\ is so weak that the FWHM measurement is meaningless.
Nevertheless, the mass determination is quite consistent with the
results of the AGN Watch program.

\paragraph{NGC 5548.} Some comments on this object appear in 
section 5. There are more 
published variability data, by far, on this object than any other, and
most of the data are very good. Only a few problems need to be pointed
out. First, the \Halpha\ region of the rms spectrum from the Wise
Observatory campaign (Netzer et al.\ 1990) has strong narrow-line
residuals that render the line-width measures unusable. The CCCD for
the associated time series is also rather ambiguous, so we do not
include these measurements in the mass determination. The \Hbeta\
profile in the rms spectrum from the fifth year of AGN Watch
monitoring (i.e., JD2448954--JD2449255) is very unusual, as shown in
Fig.\ 14; there are two peaks, a strong central peak and a weak blue
peak. The FWHM measurement is thus meaningless. Finally,
\heii\,$\lambda4686$ is prominent in the rms spectrum for the first
year of the AGN Watch campaign (JD2447509--JD2447809), but is too
heavily blended with the optical \feii\ blends to measure in the mean
spectrum; this line is not included in this part of the analysis
(cf.\ Figs.\ 4 and 5).

\paragraph{PG 1700+518.} The CCCD for this object shows two peaks, 
one of them clearly 
ascribable to correlated errors in the continuum and emission-line
light curves (Fig.\ 15).  Fortunately, the peaks are well-separated,
and we can exclude the zero-lag correlated error peak from our
analysis. This significantly increases the \Hbeta\ lag and also the
black hole mass relative to the original investigation by Kaspi et
al.\ (2000).

\paragraph{3C 390.3.} Some of the difficulties with this object 
have already been discussed in section 5. 
Part of the problem with these data seems to be just the
nature of the variability: apart from a large-scale outburst during
the early part of the monitoring campaign, the variations were very
weak. As noted earlier, the lines are very broad and blended, and the
most consistent black hole mass is obtained by using the line width
measured from the unblended side of each line and then assuming
symmetry. A more detailed attempt at deconvolution might improve this.
Finally, we note that in this case, we used
the UV continuum (at 1370\,\AA) as the driving continuum
in the cross-correlation analysis.

\paragraph{Mrk 509.} Like Mrk 110, the \heii\,$\lambda4686$ 
line is prominent in the rms 
spectrum of this object, so we therefore attempted to measure it in
each spectrum and produce a light curve. Because of blending with
optical \feii\ emission, the resulting light curve is probably not as
reliable as the \Hbeta\ light curve.

\paragraph{NGC 7469.} This object is one of a very few in which
a lag has been detected between the UV and optical continuum
variations (Wanders et al.\ 1997; Collier et al.\ 1998).
We therefore in this case use the UV continuum (at 1315\,\AA)
as the driving continuum in the cross-correlation analysis.

\section{THE MASS--LUMINOSITY RELATIONSHIP}
Our improved database can be used to investigate the
relationships between BLR radius and luminosity and
black hole mass and luminosity. The radius--luminosity
relationship is discussed in a companion paper
(Kaspi et al.\ 2004), and here we will discuss only
the mass--luminosity relationship.

We computed the optical luminosity from the flux
measurements in the original data sources that were made over
the time intervals used in this analysis. In each case,
we selected the continuum waveband closest to 5100\,\AA\
in the AGN rest frame. We corrected for Galactic
reddening using the extinction values in column (7)
of Table 1 and using the reddening curve of 
Savage \& Mathis (1979), adjusted to 
$A_V/E(B-V) = 3.2$. Luminosity distances were computed
using the redshifts given in column (5) of Table 1
and by assuming a standard flat $\Lambda$CDM cosmology
with $\Omega_{\rm B} = 0.04$,
$\Omega_{\rm DM} = 0.26$,
$\Omega_{\Lambda} = 0.70$,
and $\Ho=70$\,\Hubble. For AGNs with multiple mass
determinations, we computed the average value
of $\log \lambda L_{\lambda}$ from
each individual time series. The
optical luminosities $\log \lambda L_{\lambda}$ 
are given in column (4) of Table 8.
The corresponding uncertainties represent the 
amplitude of continuum variability during the
reverberation experiment.

In Fig.\ 16, we plot the reverberation-based masses
we have derived as a function of the mean luminosity.
This figure can be compared directly with Fig.\ 8
of Kaspi et al.\ (2000), which reveals that a much
better-defined mass--luminosity relationship results
from our improved analysis. We have also estimated
the bolometric luminosity in the same fashion as
Kaspi et al., i.e., $L_{\rm bol} \approx 
9 \lambda L_{\lambda}(5100\,{\rm \AA})$, and this scale is shown on
the top of Fig.\ 16. The diagonal lines show
the Eddington limit, and 10\% and 1\% its value.

We note that this constitutes only a preliminary
version of the mass--luminosity relationship, for
comparison with earlier work. A more exhaustive
study of this relationship is continuing.
Several comments are in order:
\begin{enumerate}
\item It is reassuring that there are no objects
above the Eddington limit (i.e., to the right of
the diagonal line), in contrast to the results of
Kaspi et al. This is because our analysis has corrected a number
of errors in earlier work, most notably removal of
residual narrow-line \Hbeta\ from many of the rms
spectra, which resulted in larger line widths and 
rather higher overall black hole masses.
The object closest to the Eddington limit is, not
surprisingly, 3C~273 (PG~1226+023).
\item The bolometric correction we have assumed
is nominal, based on a spectral energy distribution (SED)
with a strong blue bump. However, observed SEDs suggest
that a smaller ratio, e.g., 
$L_{\rm bol} \approx 5 \lambda L_{\lambda}$
(Netzer 2003), may be on average more appropriate.
Moreover, the bolometric correction we adopt
may not be appropriate for all
types of AGNs at arbitrary luminosity. The bolometric
luminosities are thus uncertain and should be treated
with some caution.
\item The optical luminosities used here have not
been corrected for the contribution of starlight from
the host galaxies. This can be a significant factor,
especially in the lower-luminosity objects. For example,
the standard aperture used in the NGC 5548 optical
monitoring program, 5\arcsecpoint0 $\times$ 7\arcsecpoint5,
admits a starlight flux of 
$F_\lambda(5100\,{\rm \AA}) \approx 3.4\times10^{-15}$\,\contunits\
(Romanishin et al.\ 1995).
Correcting for this reduces the optical luminosity
entry for NGC 5548 in Table 8 by $\sim0.24$\,dex. A program
is currently underway to determine the starlight contribution
to the optical luminosity of each of these objects.
\item Internal extinction has not been accounted
for in any way. Correction for extinction will
move objects to the right in this diagram. It is
worth noting that the object with the lowest
Eddington rate (farthest to the left of the
$0.01L_{\rm Edd}$ diagonal) is NGC 3227,
which is a rather dusty object (Pogge \& Martini 2002).
The low luminosity of this object relative to
its black hole mass may be a result of internal extinction.
\item Given the small formal error bars for
most of the objects, we believe that much of the scatter
in Fig.\ 16 is real. Moreoever, we find that the
scatter correlates with other AGN properties;
the I Zw 1-type objects generally lie
along the bottom edge of the envelope defined
by the data points. In Fig.\ 16, the NLS1s
are shown as open circles, and all of them except
NGC~4051 lie on the lower edge of the mass--luminosity
envelope. Conversely, the one object
with strongly double-peaked Balmer line 
profiles, 3C 390.3, lies along the upper
edge of the envelope. The locations of these
extreme objects on this diagram suggest that at least
some of the dispersion of the data points
correlates with Eigenvector 1, consistent with the
suggestion originated by Boroson \& Green (1992) and
reaffirmed by numerous later authors that 
Eigenvector 1 appears to be driven by Eddington ratio
$\dot{m} = \dot{M}/\dot{M}_{\rm Edd}$.
However, the {\em physical} origin of the scatter 
observed in Fig.\ 16 could
be attributable either to differences in Eddington ratio
or to inclination effects. Decreasing
inclination (i.e., from edge-on to face-on)
will translate points to the right as the 
apparent luminosity increases on account of decreased limb darkening
and downward as the rotational velocities appear to decrease.
Increasing the Eddington ratio  will
translate points in the same sense.
\item The best fit mass-luminosity relationship is found to be
\begin{equation}
\log \left( M / 10^8\,\Msun \right) =
-0.12 (\pm 0.07) + 0.79 (\pm 0.09) 
\log \left( 
\lambda L_{\lambda}(5100\,{\rm \AA})/10^{44}\,{\rm erg\,s}^{-1}
\right). 
\end{equation}
However, there is no reason to believe that there are
no selection effects operating.
Interestingly, the lower edge of the envelope seems
to parallel the lines of constant Eddingtion ratio rather well,
suggesting that the intrinsic mass--luminosity
slope may not differ signficantly from unity.

\end{enumerate}

\section{SUMMARY}
In this contribution, we have improved the
calibration of the reverberation-based AGN black hole
mass scale by decreasing random and systematic errors and by
drawing on the AGN \Msigma\ relationship to establish
a statistically accurate calibration that is tied to
other methods of black hole mass measurement. We have
undertaken a consistent reanalysis of a highly inhomogeneous
database that consists of 117 separate time series,
not including several others that were deemed to be
too poor to use, in the process accounting for a variety
of systematic effects such as time dilation on the
time lags and spectral resolution on the line widths.
Each time series is treated independently and
yields an independent estimate of the black hole mass,
thus reducing random uncertainties. Poor or suspicious
data are removed from the database, noting especially
the susceptibility of time-lag measurements to
correlated errors and other types of aliasing.
We find that the most consistent mass
measurements are obtained by using the cross-correlation
centroid \tcent\ to characterize the light-travel time 
across the BLR and by using the line dispersion
\sigline\ as measured in the rms spectrum to characterize
the velocity dispersion of the BLR. In practice, 
special care has been taken to remove residual narrow-line
contamination of the rms spectrum in cases where it is
present. The result of this analysis is a revised
AGN mass scale based on 35 reverberation-mapped AGNs
that is statistically accurate to better than a factor of three.

\acknowledgements
We are grateful for support of this work through NSF grant
AST-0205964 to The Ohio State University, through
US--Israel Binational Science Foundation grant 1999--336,
and through a grant from Israel Science Foundation
grant 232/03. This research has made use of the 
NASA/IPAC Extragalactic Database (NED) which is operated by the 
Jet Propulsion Laboratory, California Institute of Technology, 
under contract with the National Aeronautics and Space Administration.
We thank an anonymous referee for suggestions that led
to improvements in the paper.

\appendix

\section{SOME COMMENTS ON CROSS-CORRELATION
METHODOLOGY}

Welsh (1999) has suggested a number of
modifications to our error analysis procedures and we have investigated
these through additional Monte Carlo simulations similar
to those described by P98b.

Emission-line lag measurements are made by cross-correlating
the emission-line and continuum light curves. Since the
data are almost never regularly sampled, real data points
in one time series are matched with values obtained by
linear interpolation of the other time series. The cross-correlation
function (CCF) is measured twice, once interpolating in the continuum
series, and once in the emission-line series, and the final
CCF is determined by averaging these two results. The
CCF is characterized by (1) its peak value (highest value of the correlation
coefficient) \rmax, 
(2) the time delay corresponding to this
value \tpeak, and (3) the centroid \tcent\ of the peak in the CCF.
In practice, \tcent\ is evaluated using only points
with values about some threshold, usually $0.8\rmax$.

To assess the uncertainties in the determination of
\tpeak\ and \tcent, we use the model-independent Monte
Carlo FR/RSS method described by P98b.
For each Monte Carlo realization, we start with
a parent light curve of $N$ data points and from this
make $N$ independent random selections of these data points
without consideration for previous selections. The redundant
selections are then discarded, leaving a new light curve
of $M \leq N$ points; typically, the fraction of points from
the parent light curve that remain unselected in each realization
is $\sim1/e$. We refer to this process as ``random subset
selection'' (RSS), and it seems to successfully account for
uncertainties due to effects of individual data points.
Another major source of uncertainty is the uncertainty in
the measured continuum and emission-line fluxes, and these can
be significant if the amplitude of variability is not much
larger than the flux errors in individual data points.
We attempt to account for this by altering the fluxes
of the $M$ data points in each Monte Carlo realization
by random Gaussian deviates scaled by the flux error
associated with the data point. We refer to this process as
``flux randomization'' (FR). This process is repeated
for many independent realizations (usually 2000 or more
for the light curves analyzed here), and for the
CCF from each realization, \rmax, \tpeak, and \tcent\
are recorded. These are used to build up
``cross-correlation peak distributions (CCPDs)'' and
``cross-correlation centroid distributions (CCCDs)'' for
\tpeak\ and \tcent, respectively. This is done because
the distributions of these values are rarely even approximately
Gaussian (Maoz \& Netzer 1989).

Welsh (1999) points out a number of potential problems
with cross-correlation methodology, some of which can
produce biases in determination of the emission-line lags.
However, simulations that mimic as much as possible
real AGN behavior have not revealed any {\em strong}
systematic biases. Nevertheless, we caution that
cross-correlation of light curves to measure emission-line
lags is a rather crude tool, but one that seems to be
effective with the limited AGN variability data
available at present. 
Welsh has also suggested a number of
modification to the FR/RSS method, and we have extended the
Monte Carlo simulations described by P98b
to test these suggestions under what we regard as 
reasonably realistic conditions, and we describe our
results below.

\paragraph{\bf Flux Uncertainties and Redundant Selections.}
Welsh (1999) suggests an alternative strategy to FR/RSS,
namely counting redundant selections in each realization
and appropriately reducing the flux error for multiply-selected
points. Specifically, the flux uncertainty for each point
that is selected $1 \leq n \leq N$ times should be reduced
by a factor of $n^{1/2}$. Thus, rather than omitting redundant 
points, each of the $N$ selections has some real effect on
the outcome of the realization and the weighting is more
consistent with the standard bootstrap method
(e.g., Diaconis \& Efron 1983) on which the FR/RSS method
is based. We carried out detailed simulations like those
described by P98b to determine the efficacy of Welsh's
proposed scheme. We compare different error assessments
by examining the width of the CCPDs and CCCDs produced by
simulations; we presume that the algorithm that yields the
narrowest CCPDs and CCCDs (i.e., the highest precision measurements)
is the best, as long as the errors are not underestimated.
Our simulations confirm that the error estimates using
Welsh's method are superior to those of the original
FR/RSS algorithm. The uncertainties in \tpeak\ are typically
lower by $\sim8$\%, and the uncertainties in \tcent\ are
lower by $\sim3$\%. Following the procedures of P98b,
we have also carried out model-dependent Monte Carlo simulations in which
we use a known model for the transfer function in order
to verify that Welsh's algorithm does not underestimate the uncertainties.
We therefore adopt this improvement in the cross-correlation
analysis employed in this contribution.

\paragraph{Detrending the Light Curves.}
Welsh (1999) also suggests that cross-correlation results are
more accurate if the data are first ``detrended,'' i.e.,
a low-order (usually linear) polynomial is fitted
to the light curve and subtracted off prior to carrying out
the cross-correlation analysis. Our simulations support this,
but only in the case where the sampling is excellent,
i.e., long duration at high resolution (for the cases
considered by P98b, say, a 200-day experiment with observations
once per day). However, we find that under conditions of
more marginal sampling (say, a 200-day experiment with only
40 observations), which includes nearly all reverberation
data that exist, detrending either makes little difference or
can, in fact, lead to occasional gross errors in the
lag determination and, consequently, gross overestimates
of the uncertainties in the lags. 
For this reason, we elect to not detrend our
data prior to cross-correlation.

\paragraph{Peak or Centroid?}
Whether or not the cross-correlation lag is better characterized
by the peak or centroid of the CCF has been debated on many
occasions. The advantage of the centroid is that it is related
to the centroid of the transfer function and is better
defined when the CCF has a broad peak, and for these reasons,
we prefer it. Welsh's simulations, however, suggest that
the peak is a more robust measure than the centroid, quite
at odds with what was found by P98b and contrary to our
general experience. The difference seems to be attributable
to differences in the transfer functions used by Welsh 
and by P98b; P98b used only thin-shell and thick-shell transfer
functions which are flat-topped (i.e., with poorly defined peaks),
whereas Welsh used only Gaussian
transfer functions. If we repeat the simulations of P98b with
Gaussian transfer functions, we find little reason to prefer one
measure to the other. Further investigation of this issue
using real data, as described in this paper, leads us to continue
to prefer the CCF centroid, although the CCD peak is also an acceptable
way to characterize the emission-line lags.

\paragraph{Centroid Threshold.} 
Aliasing effects can lead to complex structures in CCFs; rarely
is the principal peak in the CCF isolated and well-defined.
It is therefore necessary to compute the CCF centroid only
over a restricted range, including points with values
larger than some fixed fraction of the peak value \rmax.
It is conventional to use a threshold of $0.8\rmax$ for
this computation, although in some cases where the peak is
noisy, lower thresholds are used. In some cases, the
centroid can vary significantly for different threshold
selections (Koratkar \& Gaskell 1991), though this appears
to be less of a problem with well-sampled data
(e.g., Dietrich et al.\ 1993).
Simulations based on those of P98b, however, show that
a threshold value of $0.8\rmax$ is generally a good choice.
With lower thresholds, we find that the CCCD is significantly
broadened. For a threshold of $0.5\rmax$, for example, we find 
that the width of the CCCD increases by 10--20\%, depending
somewhat on the details of the transfer function (i.e.,
more sharply peaked transfer functions give results less
sensitive to the selected threshold value, as one might expect).
We will therefore continue to use a threshold
of $0.8\rmax$, unless otherwise noted.


\begin{figure}
\plotone{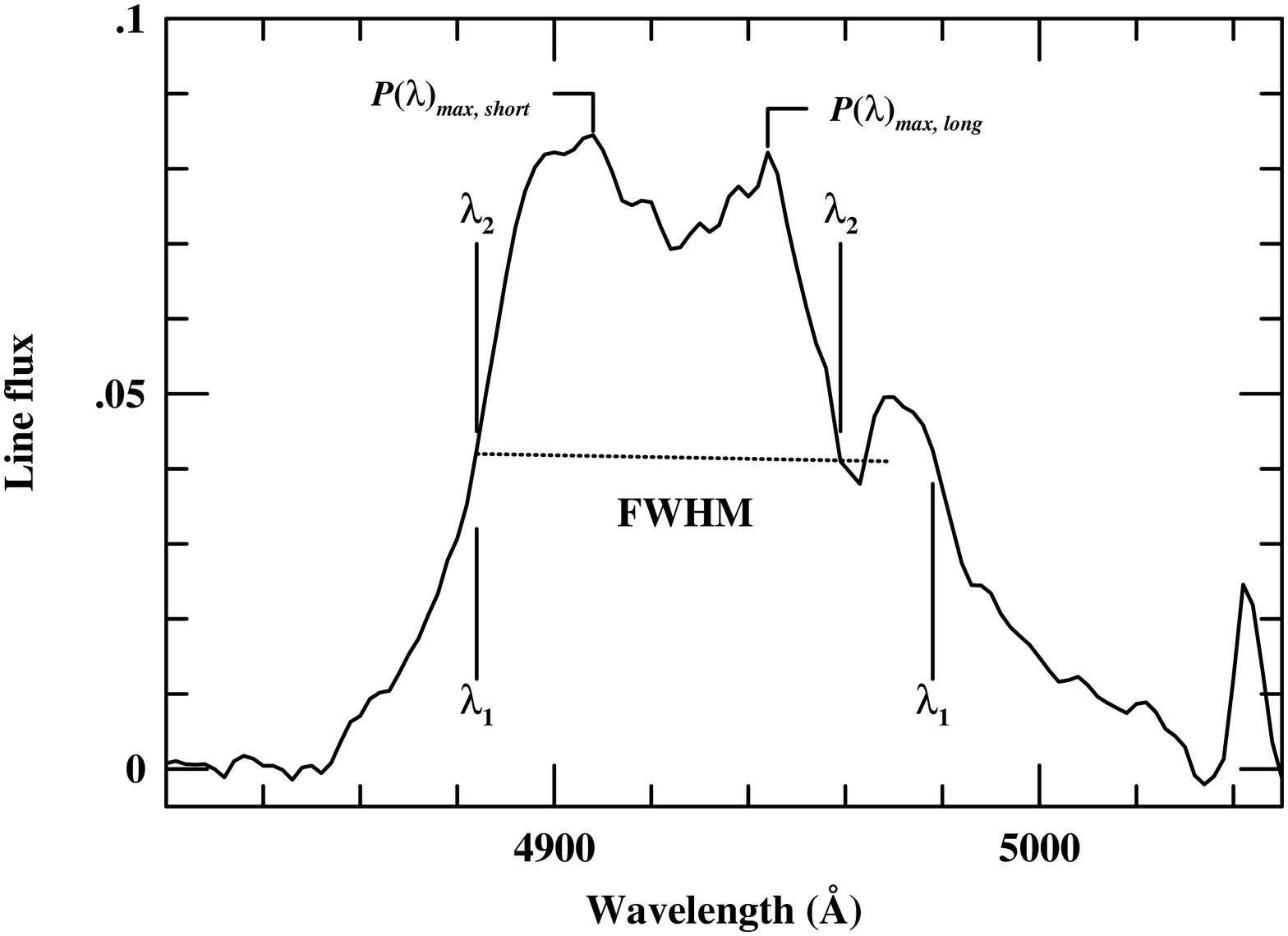}
\caption{Measurement of FWHM for double-peaked profiles.
A continuum is interpolated underneath the line profile
and is subtracted from the data.
Peak fluxes are identified on the short-wavelength and
long-wavelength peaks, $P(\lambda)_{max, short}$
and $P(\lambda)_{max, long}$, respectively.
On each side of the line, wavelengths corresponding to
the half-maximum fluxes 
$0.5P(\lambda)_{max, short}$
and $0.5P(\lambda)_{max, long}$ are found
moving upward from the continuum at $\lambda_1$ and
downward from the peak at $\lambda_2$. In this example,
$\lambda_1 = \lambda_2$ on the short-wavelength side only.
The nominal position of the half-maximum point is taken to
be the average of $\lambda_1$ and $\lambda_2$.}
\end{figure}

\begin{figure}
\plotone{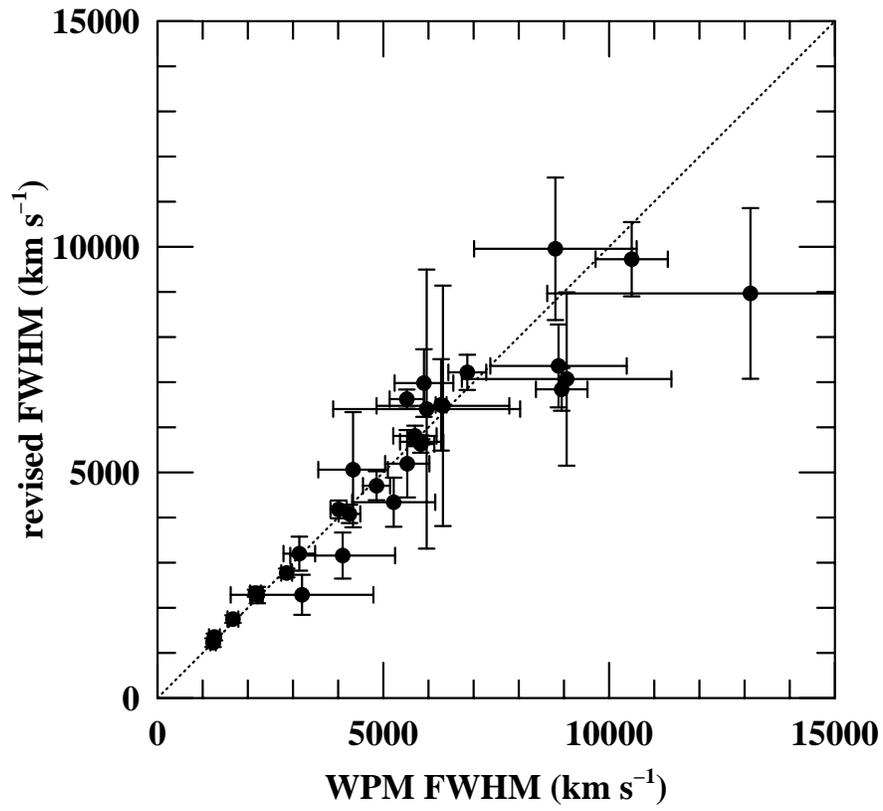}
\caption{Comparison of FWHM measurements and uncertainties from
this paper with those from WPM. For the sake of meaningful
comparison, the new measurements here have {\em not} been
corrected for spectral resolution, as have all other 
line-width measurements in this paper.}
\end{figure}

\begin{figure}
\plotone{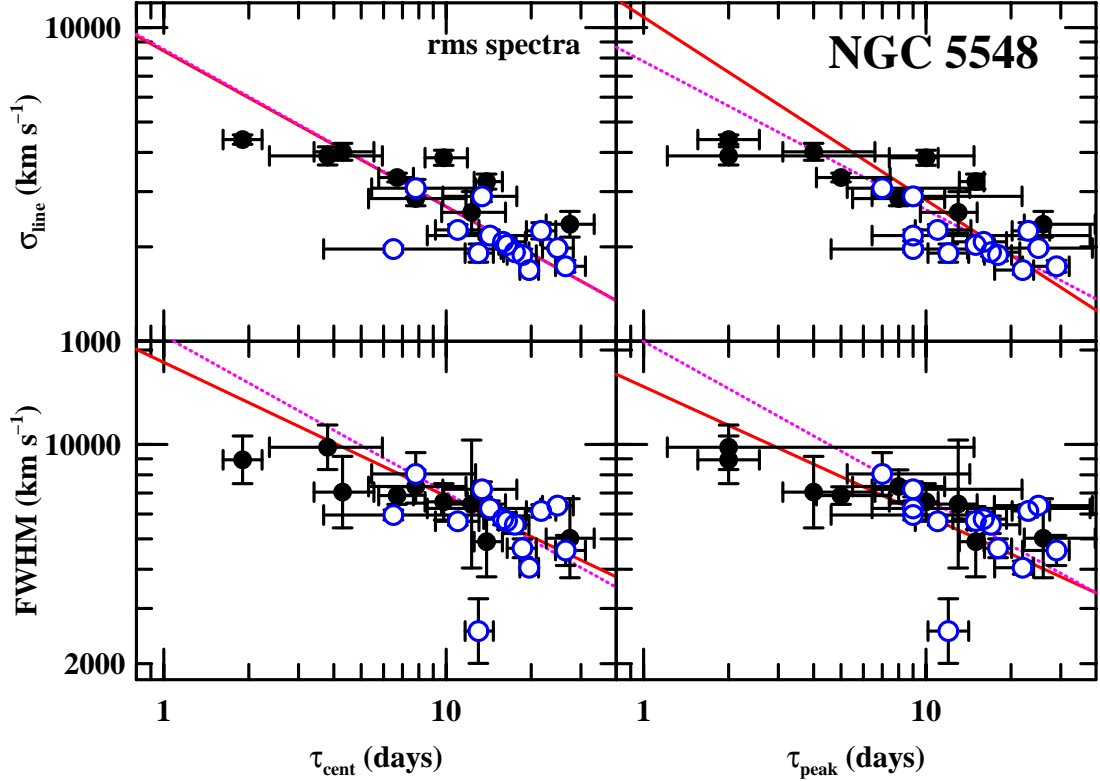}
\caption{Line widths versus time lags for emission lines in the
rms spectra of NGC 5548. The top row shows the line dispersion
\sigline\ as a line width measure, and the bottom row shows
FWHM. The left column shows the CCF centroid \tcent\ as the
time-lag measure and the right column shows the location
of the CCF peak \tpeak. The solid line is the best fit to
the data, and the dotted line is a forced fit to
slope $-1/2$, the virial slope. The fit parameters are 
summarized in Table 7. The open circles are measurements
of \Hbeta\ for 14 different years. The filled circles
represent all of the other lines.}
\end{figure}

\begin{figure}
\plotone{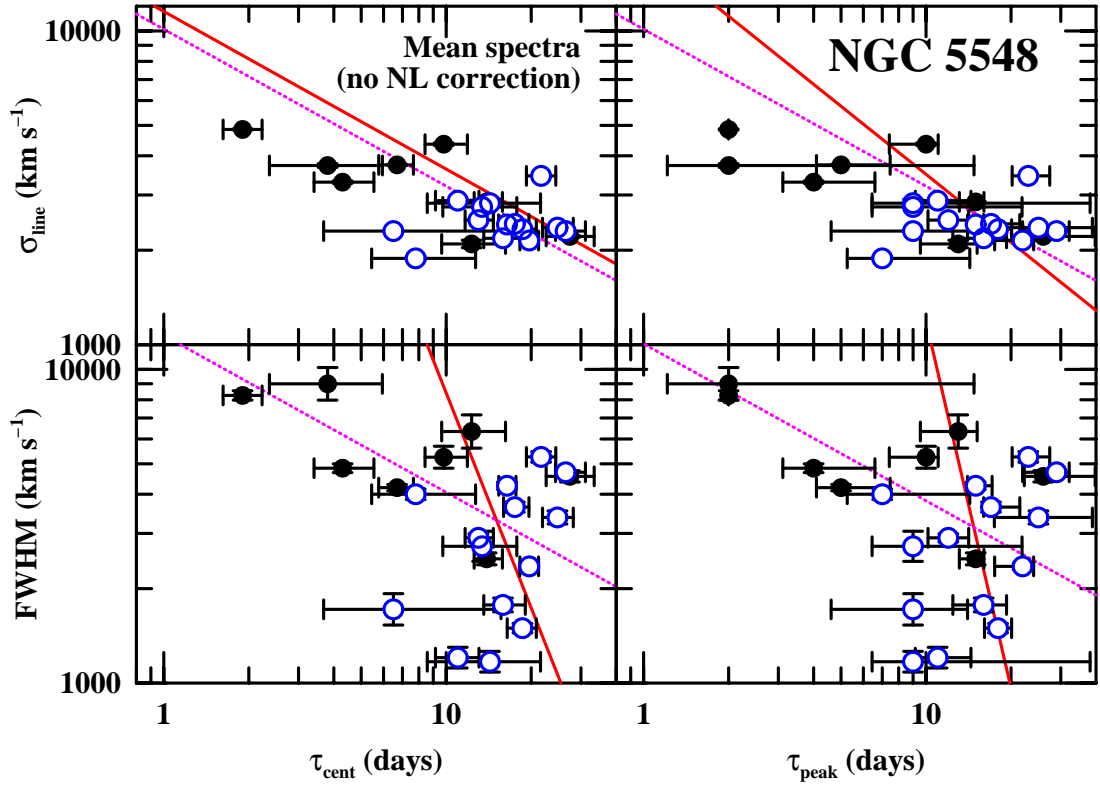}
\caption{Line width versus time lags for emission lines in the
mean spectra of NGC 5548. The data are plotted as in Fig.\ 3.
Note the dramatic change in the scatter, especially for the
\Hbeta\ line, by using the mean rather than the rms line profile
for the line-width measurement.}
\end{figure}

\begin{figure}
\plotone{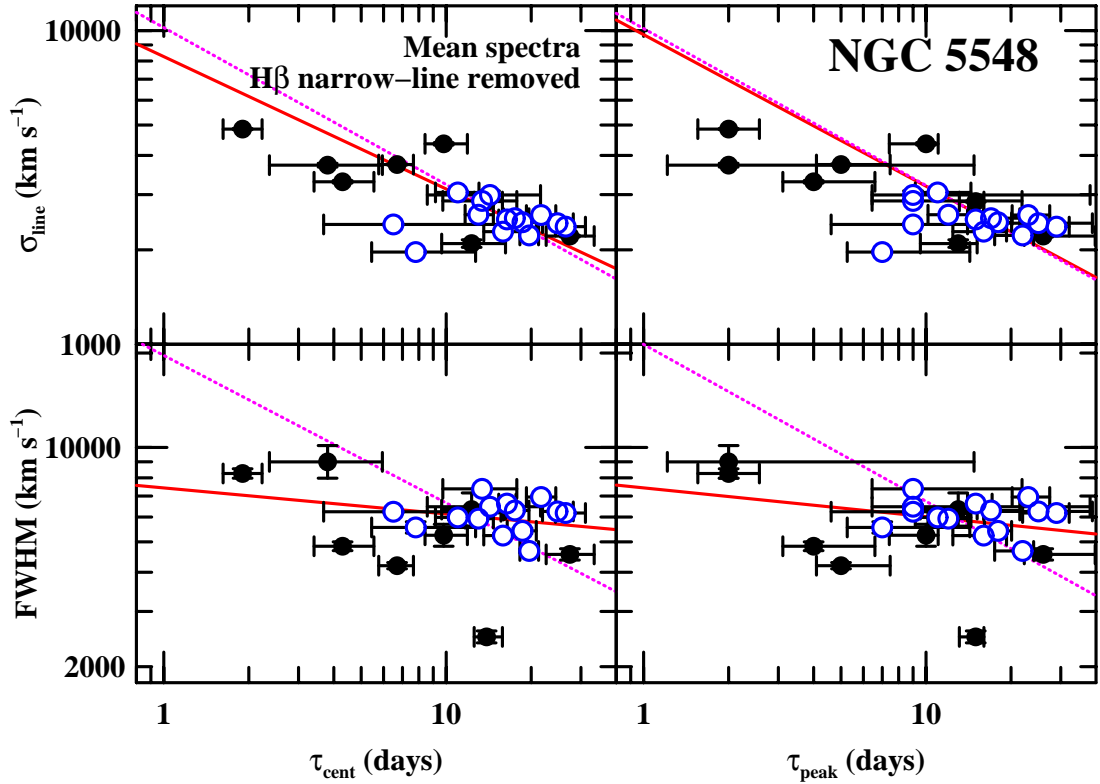}
\caption{Line widths versus time lags for emission lines in the
mean spectra of NGC 5548, but with the narrow component
of \Hbeta\ removed.  The data are plotted as in Figs.\ 3 and 4.
Removing the narrow-component of \Hbeta\ greatly reduces the
scatter around the virial relationship, but the scatter
is still much more pronounced than in Fig.\ 3, where the
line-widths are measured in the rms profiles. The measurement
uncertainties on \sigline\ are too small to show up on this
diagram. Note that the fits to the data in the lower panel
are unweighted fits on account of the large number of outliers.}
\end{figure}

\begin{figure}
\plotone{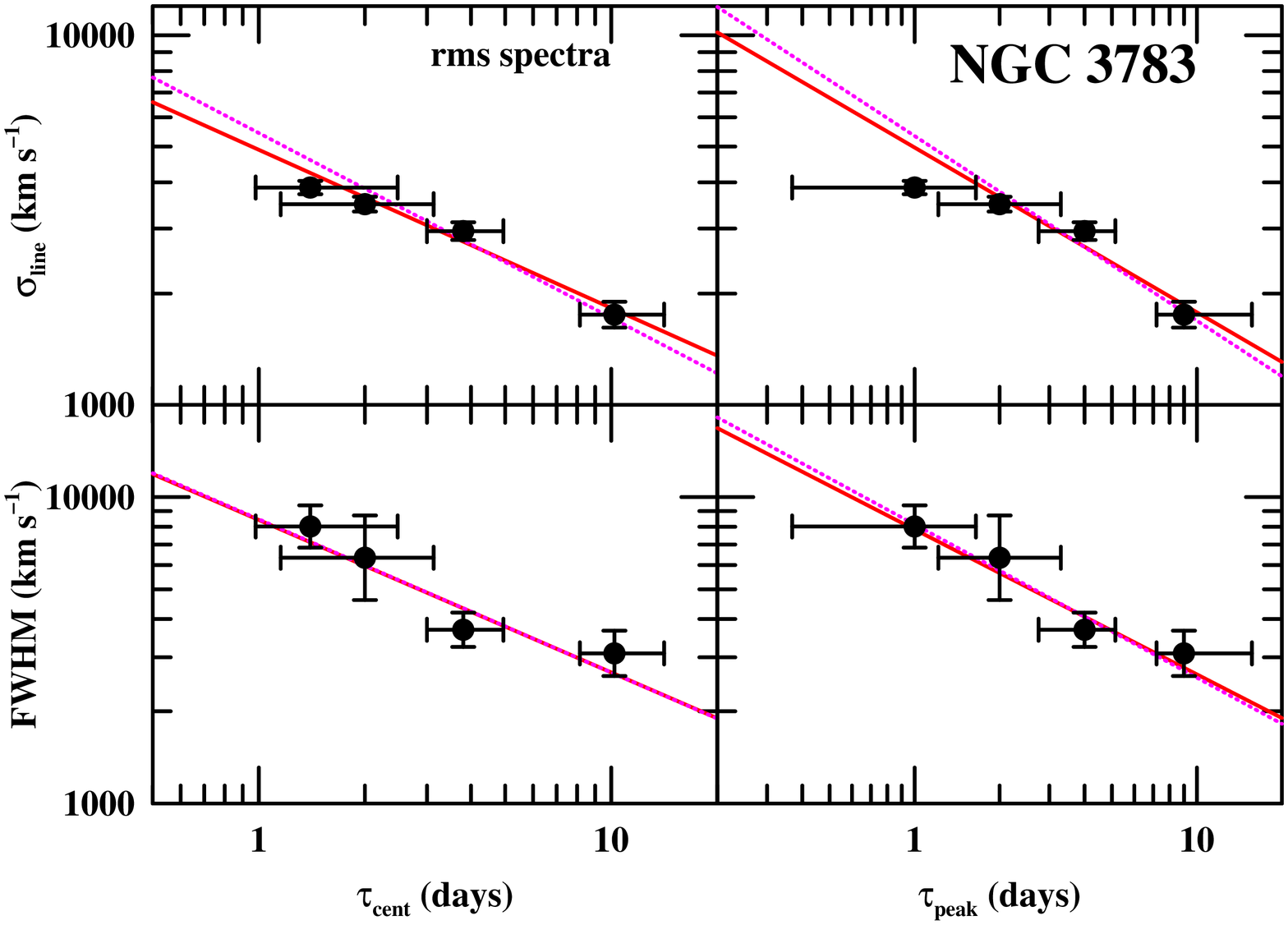}
\caption{Line widths versus time lags for emission lines in the
rms spectra of NGC 3783. The data are plotted as in Fig.\ 3.}
\end{figure}

\begin{figure}
\plotone{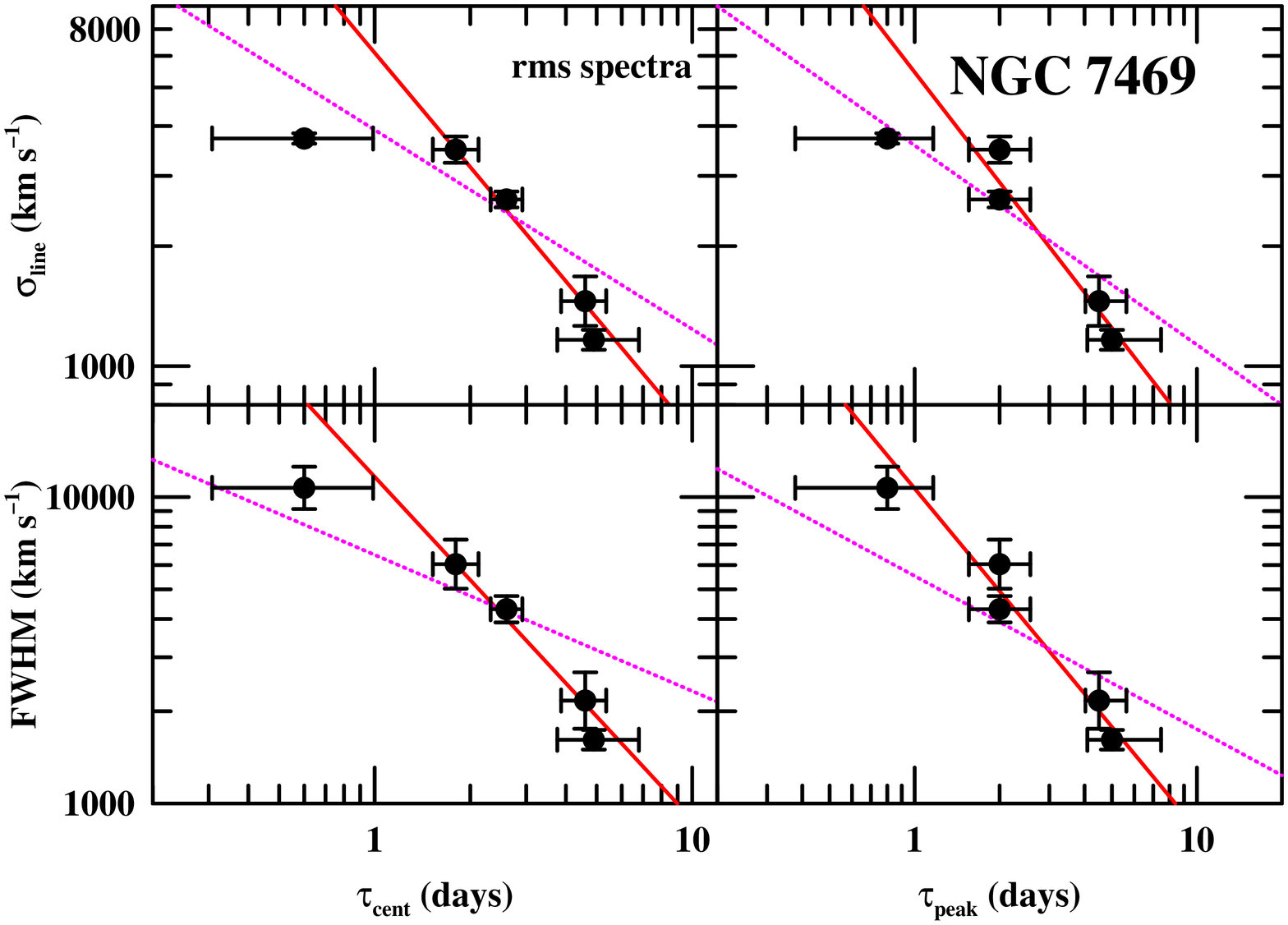}
\caption{Line widths versus time lags for emission lines in the
rms spectra of NGC 7469. The data are plotted as in Fig.\ 3.}
\end{figure}

\begin{figure}
\plotone{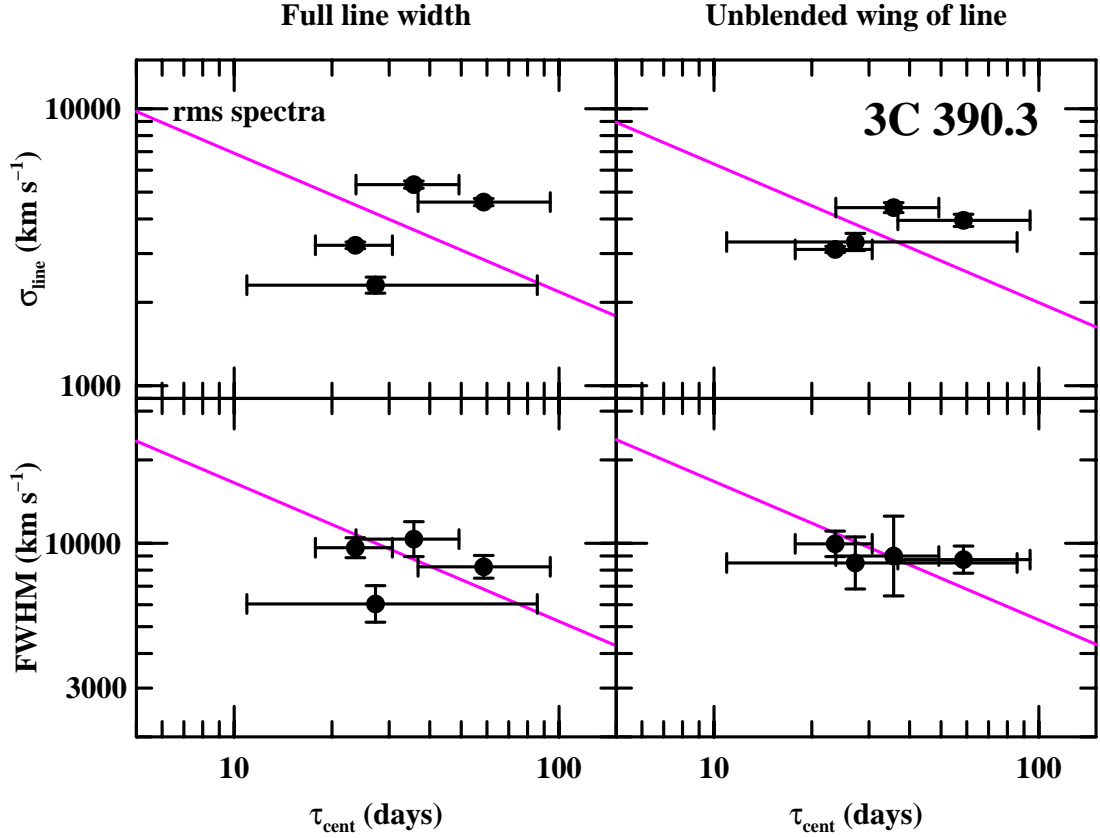}
\caption{Line width versus CCF centroid \tcent\ for emission lines in the
rms spectra of 3C 390.3. The top row shows the line dispersion
\sigline\ used as the line-width measure, and the bottom row shows
FWHM. The left-hand column shows our standard line-width measurement,
using the whole line, but with no effort to account for blending.
In the right-hand column, we attempt to account for blending by
measuring the unblended half of each line. In any case,
the large uncertainties in the time lags preclude a critical
test of a virial relationship.}
\end{figure}

\begin{figure}
\plotone{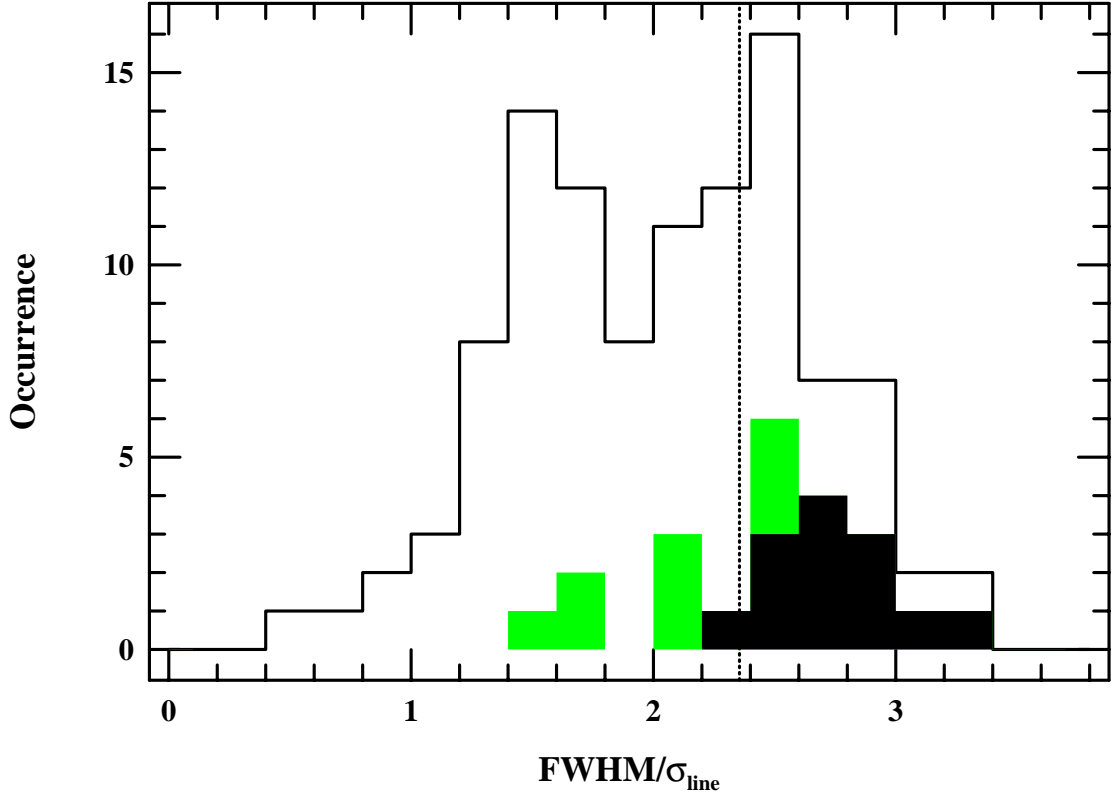}
\caption{Distribution of values of ${\rm FWHM}/\sigline$ for
all lines in the rms spectra used
in this analysis (i.e., columns (5) and (6) of Table 6),
excluding highly uncertain values (preceeded with a colon in Table 6).
The mean and standard deviation for the total distribution are
$2.03\pm0.59$.
The black area shows the distribution for multiple measurements of
\Hbeta\ in NGC 5548 (the mean and standard deviation for this subset are
$2.73\pm0.24$). The gray area shows the distribution for
other lines in NGC 5548 (mean and standard deviation for all
lines in NGC 5548, including \Hbeta, are $2.45\pm0.44$.
The vertical dotted line is at
${\rm FWHM}/\sigline = 2.355$, which is appropriate for
a Gaussian line profile. Values smaller than this indicate lines
that have weaker cores and strong wings relative to a Gaussian.}
\end{figure}

\begin{figure}
\plotone{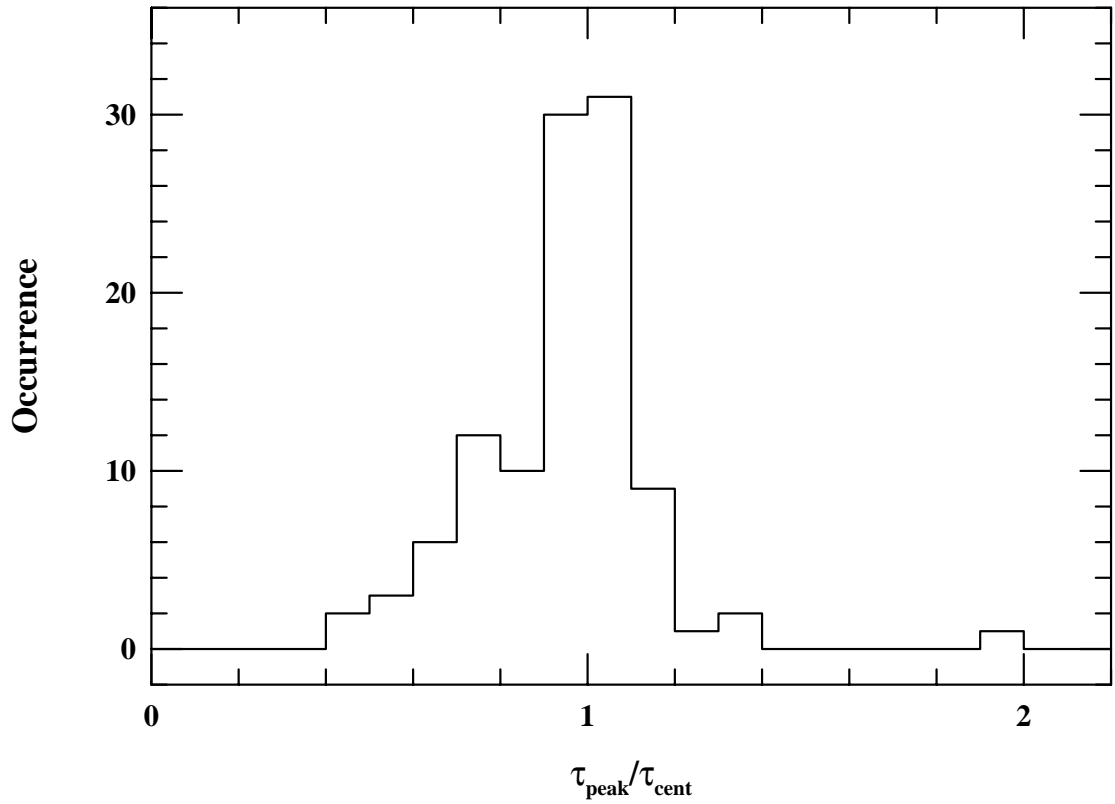}
\caption{Distribution of values of $\tpeak/\tcent$ for
all lines in this analysis (i.e., columns (3) and (4) of Table 6),
excluding highly uncertain values (preceeded with a colon in Table 6).
The mean and standard deviation for this distribution are
$0.95\pm0.20$.}
\end{figure}

\begin{figure}
\plotone{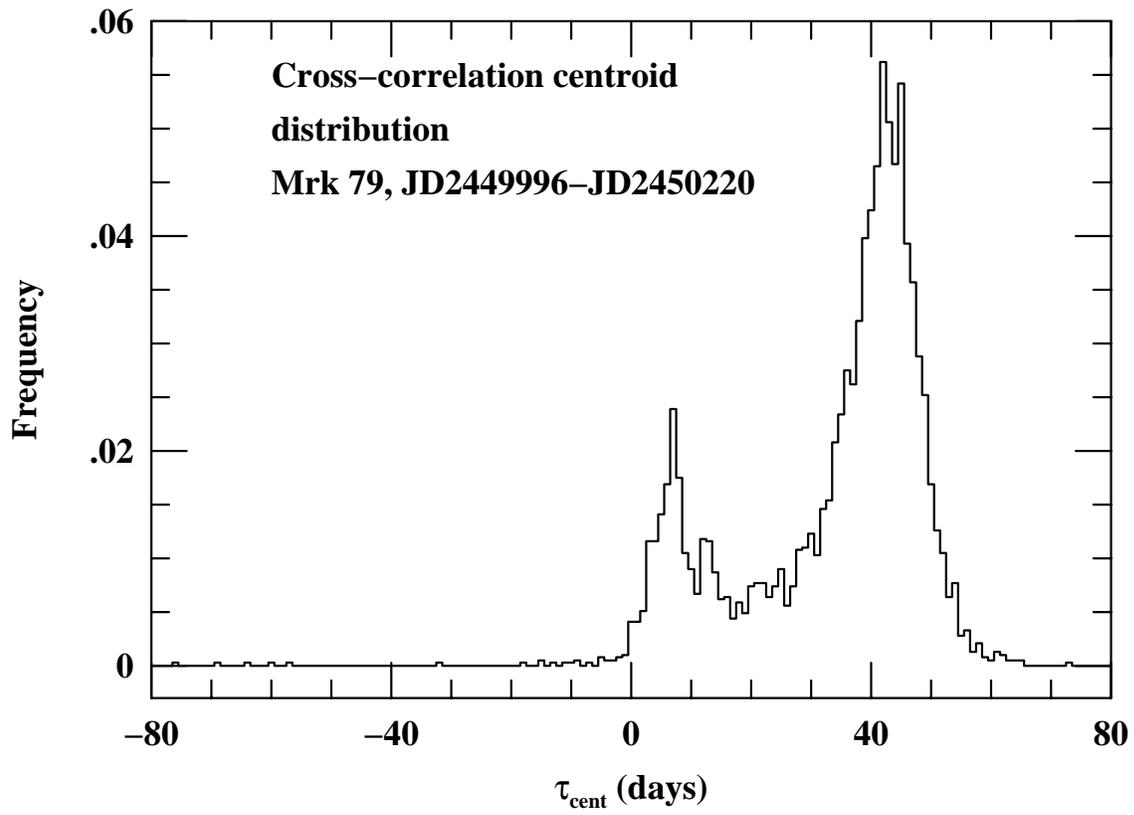}
\caption{Cross-correlation centroid distribution for the
continuum--\Hbeta\ cross-correlation for Mrk 79 during the
period JD2449996 to JD2450220. It is not obvious which peak
corresponds to the correct lag.}
\end{figure}

\begin{figure}
\epsscale{0.8}
\plotone{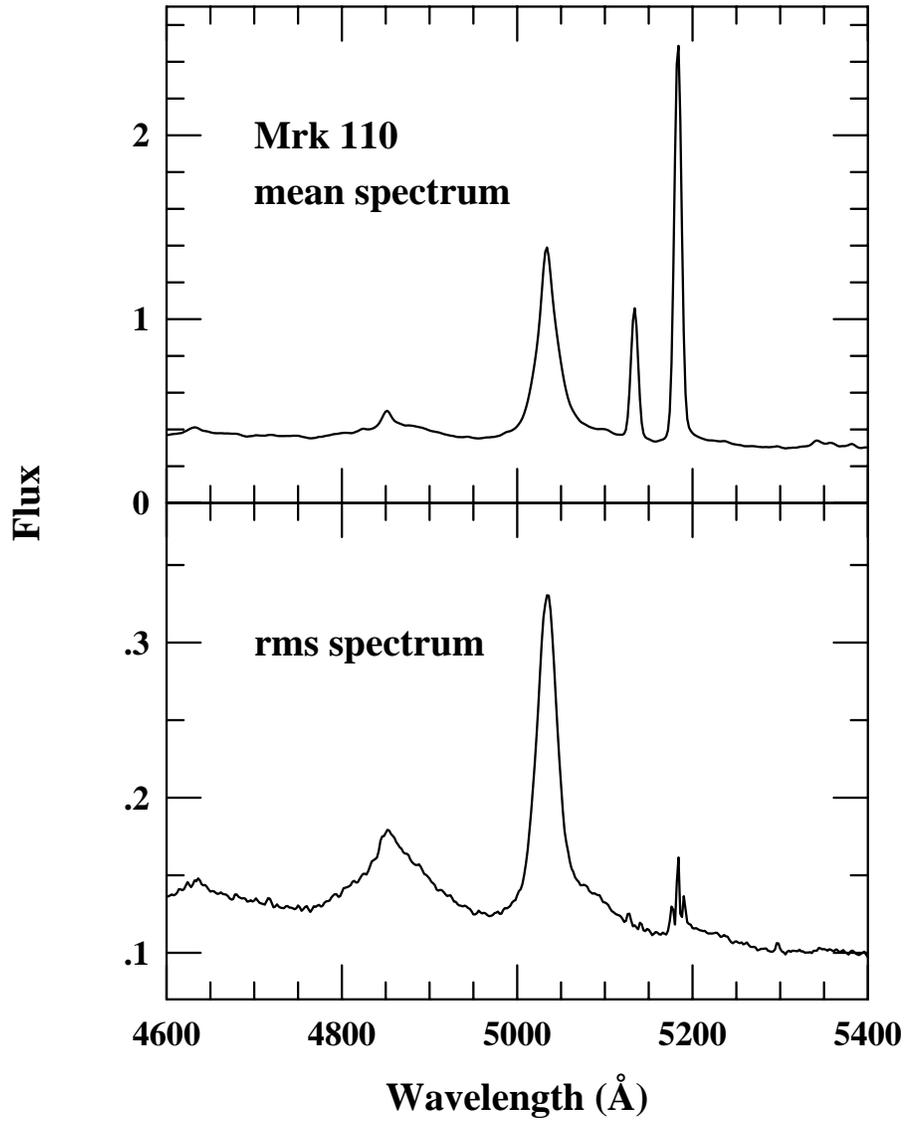}
\caption{Mean (top panel) and rms (bottom panel) spectra of
Mrk 110. The rms spectrum shows clearly the very broad
\heii\,$\lambda4686$ and broad \Hbeta. The narrow
[\oiii]\,$\lambda\lambda4959$, 5007 lines appear only
in the mean spectrum, except for weak residuals in
\o5007\ that appear in the rms spectrum.}
\end{figure}

\begin{figure}
\epsscale{1.0}
\plotone{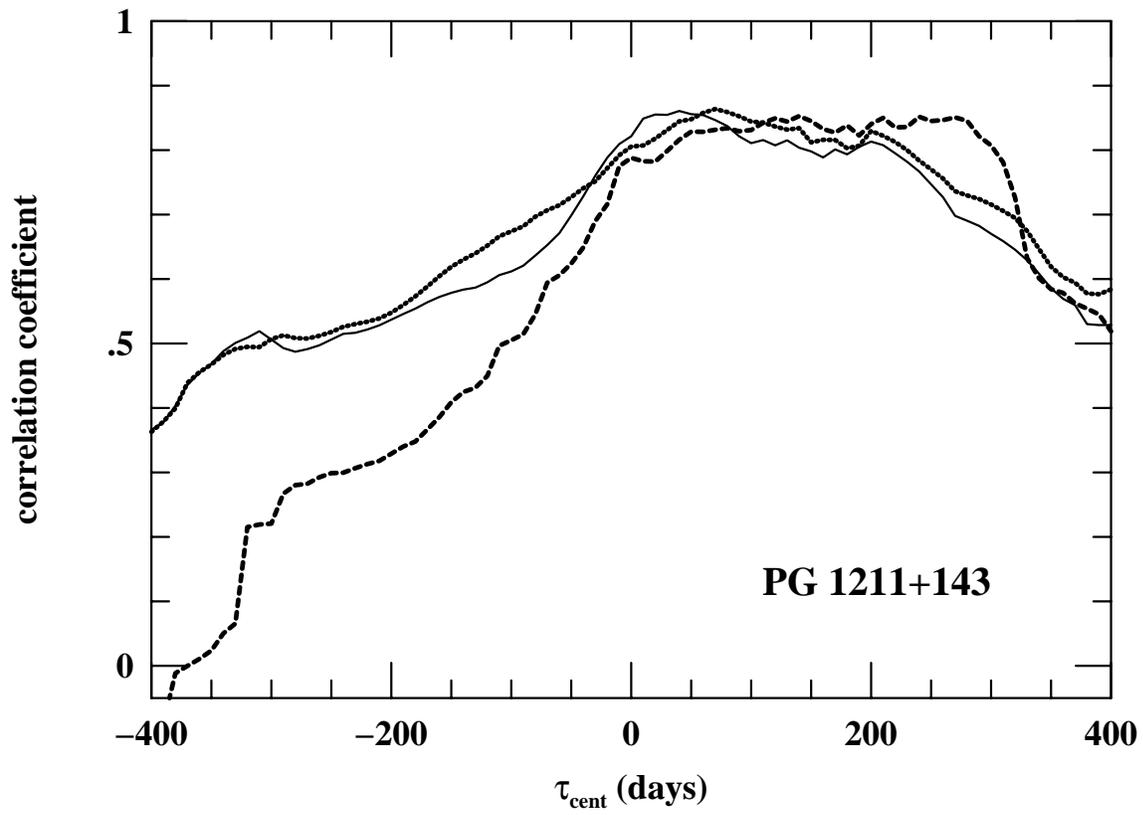}
\caption{Cross-correlation functions for PG 1211+143.
The dotted line is the \Halpha\ CCF, the solid line
is the \Hbeta\ CCF, and the dashed line is the
\Hgamma\ CCF. The centroids and peaks are poorly
defined because the CCFs are broad and flat-topped.}
\end{figure}

\begin{figure}
\epsscale{0.8}
\plotone{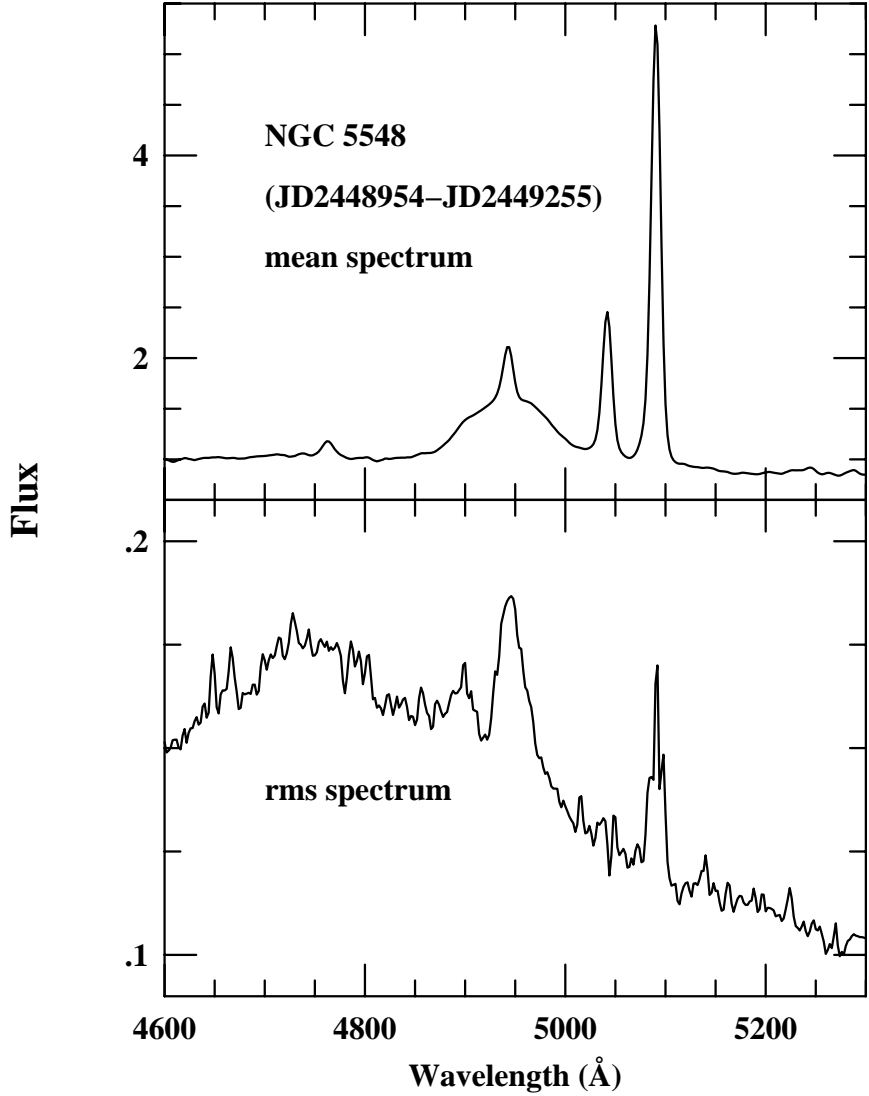}
\caption{Mean (top panel) and rms (bottom panel) \Hbeta-region spectra of
NGC 5548 during the period JD2448954 to JD2449255, the fifth
year of monitoring by the International AGN Watch (1993). 
The \Hbeta\ profile in the rms spectrum is double-peaked; there
is a strong peak at line center and another in the shortward
wing of the line.}
\end{figure}

\begin{figure}
\epsscale{1.0}
\plotone{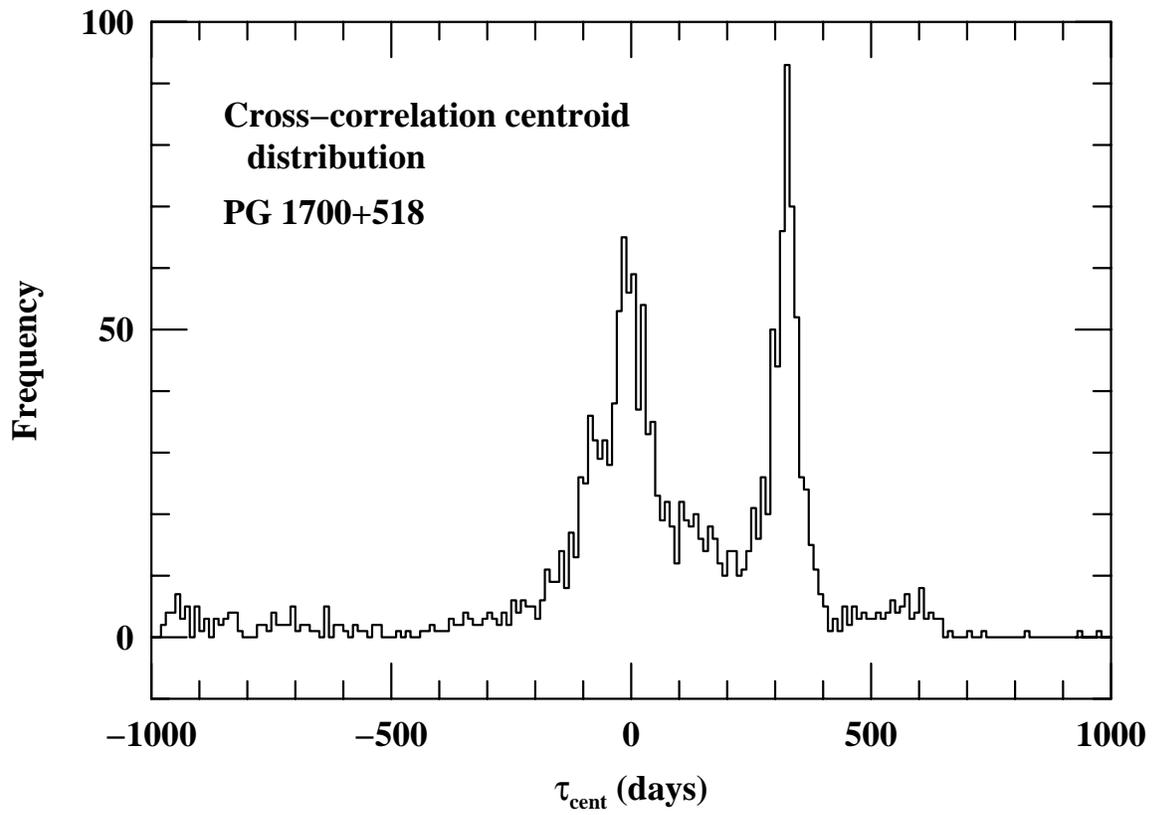}
\caption{Cross-correlation centroid distribution for the
continuum--\Hbeta\ cross-correlation for PG 1700+518.
The peak at zero lag is clearly ascribable to correlated
error, so it can be rejected.}
\end{figure}

\begin{figure}
\plotone{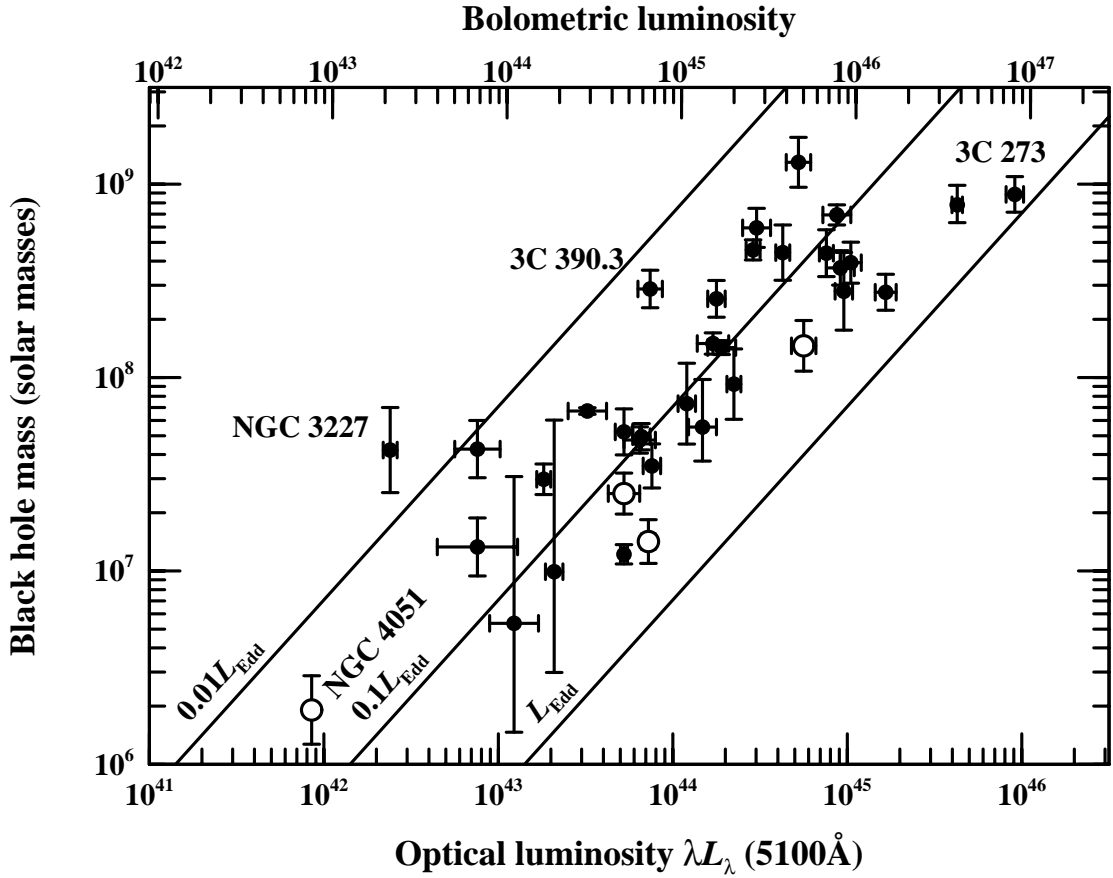}
\caption{Black hole mass vs.\ luminosity for 35 reverberation
mapped AGNs. The luminosity scale on the lower x-axis is
$\log \lambda L_{\lambda}$ in units of ergs s$^{-1}$. The
upper x-axis shows the bolometric luminosity assuming that
$L_{\rm bol} \approx 9\lambda L_{\lambda}$. The diagonal
lines show the Eddington limit $L_{\rm Edd}$,
$0.1 L_{\rm Edd}$, and $0.01L_{\rm Edd}$. The open circles
represent NLS1s. Other labeled points are discussed in
the text.}
\end{figure}

\clearpage

\tabletypesize{\scriptsize}

%
%
\begin{deluxetable}{lcllcccll}
\tablewidth{0pt}
\tablecaption{Object List}
\tablehead{
\colhead{} &
\colhead{} &
\colhead{$\alpha_{2000}$} &
\colhead{$\delta_{2000}$} &
\colhead{$z$} &
\colhead{$V$} &
\colhead{$A_B$} &
\colhead{V\'{e}ron-Cetty \& V\'{e}ron} &
\colhead{Alternative} \\
\colhead{Objects} &
\colhead{References\tablenotemark{1}} &
\colhead{(hr min sec)} &
\colhead{($^{\rm o}\ '\ ''$)} &
\colhead{ } &
\colhead{(mag)}&
\colhead{(mag)}&
\colhead{Catalog Name} &
\colhead{Name} \\
\colhead{(1)} &
\colhead{(2)} &
\colhead{(3)} &
\colhead{(4)} &
\colhead{(5)} &
\colhead{(6)} &
\colhead{(7)} &
\colhead{(8)} &
\colhead{(9)} 
}
\startdata
Mrk 335 &1& $00\ 06\ 19.521$ & $+20\  12\  10.49$ & 0.02578 & 13.7 & 0.153 & 
	J000619.5$+$201211 & PG 0003$+$199\\
PG 0026$+$129 &2& $00\ 29\ 13.600$ & $+13\ 16\ 03.00$ & 0.14200 & 15.4 
        & 0.307 & J002913.8$+$131605\\
PG 0052$+$251&2 & $00\ 54\ 52.100$ & $+25\ 25\ 38.00$ & 0.15500 & 15.4 
        & 0.205 & J005452.2$+$252539\\
Fairall 9 &3,4& $01\ 23\ 45.780$ & $-58\ 48\ 20.50$ & 0.04702 & 13.5 & 0.116 & 
	J012345.8$-$584821\\
Mrk 590 &1& $02\ 14\ 33.562$ & $-00\ 46\ 00.09$ & 0.02638 & 13.8 & 0.161 & 
	J021433.6$-$004600 & NGC 863\\
3C 120 &1& $04\ 33\ 11.095$ & $+05\ 21\ 15.62$ & 0.03301 & 14.1 & 1.283 & 
	J043311.1$+$052115 & Mrk  1506\\
Akn 120&1 & $05\ 16\ 11.421$ & $-00\ 08\ 59.38$ & 0.03230 & 14.1 & 0.554 & 
	J051611.4$-$000900 & Mrk  1095\\
Mrk 79&1& $07\ 42\ 32.797$ & $+49\ 48\ 34.75$ & 0.02219 & 13.9 & 0.305 & 
	J074232.8$+$494835\\
\enddata
\tablenotetext{1}{References: 1: Peterson et al.\ 1998a;
2: Kaspi et al.\ 2000;
3: Santos-Lle\'{o} et al.\ 1997;
4: Rodr\'{\i}guez-Pascual et al.\ 1997;
5: Salamanca et al.\ 1994;
6: Onken et al.\ 2003;
7: Winge et al.\ 1995;
8: Wanders et al.\ 1993;
9: Stirpe et al.\ 1994;
10: Onken \& Peterson 2002;
11: Reichert et al.\ 1994;
12: Peterson et al.\ 2000;
13: Kaspi et al.\ 1996;
14: Maoz et al.\ 1991;
15: Dietrich et al.\ 1994;
16: Winge et al.\ 1996;
17: Santos-Lle\'{o} et al.\ 2001;
18: Maoz et al.\ 1990;
19: Peterson et al.\ 2002 and references therein;
20: Dietrich et al.\ 1993;
21: Clavel et al.\ 1991;
22: Korista et al.\ 1995;
23: Netzer et al.\ 1990;
24: Dietrich et al.\ 1998;
25: O'Brien et al.\ 1998;
26: Collier et al.\ 1998;
27: Wanders et al.\ 1997.}
\end{deluxetable}

\tabletypesize{\footnotesize}
%
%
\begin{deluxetable}{lllccccc}
\tablewidth{0pt}
\tablecaption{Wavelength Windows for Line Width Measurements}
\tablehead{
\colhead{} &
\colhead{} &
\colhead{Data} &
\colhead{Res.} &
\colhead{Julian Dates} &
\colhead{Continuum} &
\colhead{Continuum} &
\colhead{Line}  \\
\colhead{Object} &
\colhead{Ref.\tablenotemark{1}} &
\colhead{Set} &
\colhead{(\AA)} &
\colhead{(-2400000)} &
\colhead{Window (\AA)} &
\colhead{Window (\AA)} &
\colhead{Limits (\AA)} \\
\colhead{(1)} &
\colhead{(2)} &
\colhead{(3)} &
\colhead{(4)} &
\colhead{(5)} &
\colhead{(6)} &
\colhead{(7)} &
\colhead{(8)} \\ 
}
\startdata
Mrk 335		& 1  & H$\beta$	& 9.6	& 49156--49338	& 4900--4910& 5095--5105& 4930--5030 \\ 
		& 1  & H$\beta$	& 9.6	& 49889--50118	& 4910--4920& 5100--5110& 4930--5030 \\ 	
PG 0026+129	& 2  & H$\alpha$	& 12.5	& 48836--51084	& 7290--7300& 7700--7740& 7304--7728 \\ 	
		& 2  & H$\beta$	& 12.5	& 48545--51084	& 5442--5450& 5820--5840& 5460--5700 \\ 	
PG 0052+251	& 2  & H$\alpha$	& 12.5	& 48837--51084	& 7028--7108& 7725--7800& 7398--7724 \\ 	
		& 2  & H$\beta$	& 12.5	& 48461--51084	& 5418--5488& 5908--5978& 5518--5698 \\ 	
		& 2  & H$\gamma$	& 12.5	& 48461--51084	& 4864--4914& 5418--5488& 4926--5118 \\ 	
Fairall 9	& 3  & H$\beta$	& 11.2	& 50473--50665	& 4936--4946& 5295--5323& 4946--5295 \\
\enddata
\tablenotetext{1}{References as in Table 1.}
\end{deluxetable}

%
%
\begin{deluxetable}{lc}
\tablewidth{0pt}
\tablecaption{[O\,{\sc iii}] $\lambda5007$ Line Widths}
\tablehead{
\colhead{} &
\colhead{FWHM([O\,{\sc iii}]\,$\lambda5007$)} \\
\colhead{Object} &
\colhead{(km s$^{-1}$)} \\
\colhead{(1)} &
\colhead{(2)} 
}
\startdata
Mrk 335 & 280  \\
Fairall 9 & 425  \\
Mrk 590 & 400 \\
Akn 120 & 490 \\
Mrk 79 & 350 \\
NGC 3227 & 485 \\
NGC 3516 & 250 \\
NGC 3783 & 230 \\
NGC 4051 & 190 \\
NGC 4151 & 425 \\
NGC 4593 & 255 \\
IC 4329A & 550 \\
Mrk 279 & 580  \\
NGC 5548 & 410 \\
Mrk 817 & 330  \\
Mrk 509 & 520  \\
NGC 7469 & 360 \\
\enddata
\end{deluxetable}

%
%
\begin{deluxetable}{lcc}
\tablewidth{0pt}
\tablecaption{Strength of Narrow-H$\beta$ Component}
\tablehead{
\colhead{Object} &
\colhead{$F({\rm H}\beta_{\rm narrow})/F
\mbox{([O\,{\sc iii}]\,}\lambda5007$)} &
\colhead{Ref.\tablenotemark{1}} \\
\colhead{(1)} &
\colhead{(2)} &
\colhead{(3)} 
}
\startdata
PG 0026$+$129 & 0.108 & 1 \\
PG 0052$+$251 & 0.120 & 1 \\
Fairall 9 & 0.121 & 1\\
PG 0953$+$414 & 0.147 & 1 \\
NGC 3227 & 0.088 & 2 \\
NGC 4151 & 0.084 & 2 \\
PG 1229$+$204& 0.080 & 2 \\
PG 1307$+$085& 0.094 &1 \\
IC 4329A & 0.136 & 1\\
Mrk 279  & 0.215 & 1 \\
PG 1351$+$640 & 0.330 & 1 \\
PG 1411$+$442 & 0.213 & 1 \\
NGC 5548 & 0.110  & 2 \\
PG 1426$+$015 & 0.120 & 2 \\
PG 1613$+$658 & 0.075 & 1 \\ 
PG 1617$+$175 & 0.192 & 1 \\
PG 1704$+$608 & 0.110 & 2 \\
PG 2130$+$099 & 0.306 & 1 \\
NGC 7469 & 0.181 & 1 \\
\enddata
\tablenotetext{1}{References: 1: Marziani et al.\ 2002; 2: This work.}
\end{deluxetable}

%
%
\begin{deluxetable}{lccccccc}
\tablewidth{0pt}
\tablecaption{Time-Series Analysis}
\tablehead{
\colhead{ } &
\colhead{Data} &
\colhead{Julian Dates} &
\colhead{} &
\colhead{$F_{\rm var}$} &
\colhead{$F_{\rm var}$} &
\colhead{$\tau_{\rm cent}$ } &
\colhead{$\tau_{\rm peak}$ } \\
\colhead{Object} &
\colhead{Set} &
\colhead{(-2400000)} &
\colhead{$r_{\rm max}$} &
\colhead{(cont)} &
\colhead{(line)} &
\colhead{(days)} &
\colhead{(days)}\\
\colhead{(1)} &
\colhead{(2)} &
\colhead{(3)} &
\colhead{(4)} &
\colhead{(5)} &
\colhead{(6)} &
\colhead{(7)} &
\colhead{(8)} \\ 
}
\startdata
Mrk 335		& H$\beta$	& 49156--49338	& $0.829 \pm 0.066$ & 	0.066	& 0.041	& $17.3	^{+4.9	}_{-4.3	}$ & $18^{+5	}_{-6    }$ \\ 
		& H$\beta$	& 49889--50118	& $0.714 \pm 0.094$ & 	0.049	& 0.015	& $12.8	^{+6.8	}_{-5.7	}$ & $13^{+9	}_{-7    }$ \\ 	
PG 0026+129	& H$\alpha$	& 48836--51084		& $0.796 \pm 0.047$ & 	0.115	& 0.092	& $112.1^{+32.3	}_{-29.1}$ & $108^{+44	}_{-36   }$ \\ 
		& H$\beta$	& 48545--51084		& $0.814 \pm 0.047$ & 	0.173	& 0.079	& $126.8^{+27.5	}_{-32.3}$ & $105^{+37	}_{-22   }$ \\ 
PG 0052+251	& H$\alpha$	& 48837--51084		& $0.646 \pm 0.063$ & 	0.192	& 0.101	& $189.1^{+67.6	}_{-44.2}$ & $194^{+95	}_{-89   }$ \\ 
		& H$\beta$	& 48461--51084		& $0.748 \pm 0.066$ & 	0.199	& 0.114	& $103.7^{+28.3	}_{-27.8}$ & $78^{+19	}_{-14   }$ \\ 
		& H$\gamma$	& 48461--51084		& $0.672 \pm 0.067$ & 	0.199	& 0.199	& $94.2	^{+22.5	}_{-18.4}$ & $75^{+12	}_{-7    }$ \\ 
Fairall 9	& H$\beta$	& 50473--50665		& $0.715 \pm 0.067$ & 	0.328	& 0.043	& $18.2	^{+3.3	}_{-4.5	}$ & $19^{+3	}_{-4    }$ \\
\enddata
\end{deluxetable}

\tabletypesize{\footnotesize}
%
%
\begin{deluxetable}{lcccccc}
\tablewidth{0pt}
\tablecaption{Rest-Frame Lags, Line Widths, and Virial Products}
\tablehead{
\colhead{} &
\colhead{} &
\colhead{$\tau_{\rm cent}$} &
\colhead{$\tau_{\rm peak}$}&
\colhead{$\sigma_{\rm line}$} &
\colhead{FWHM} &
\colhead{$c\tau_{\rm cent}\sigma_{\rm line}^2/G$} \\
\colhead{Object} &
\colhead{Line} &
\colhead{(days)} &
\colhead{(days)}&
\colhead{(km\,s$^{-1}$)} &
\colhead{(km\,s$^{-1}$)} &
\colhead{($10^6\,M_{\odot}$)} \\
\colhead{(1)} &
\colhead{(2)} &
\colhead{(3)} &
\colhead{(4)} &
\colhead{(5)} &
\colhead{(6)} &
\colhead{(7)} \\
}
\startdata
 Mrk 335    &H$\beta$ & $  16.8^{+   4.8}_{  -4.2}$ & $ 18^{+  5}_{ -6}$ &
$  917 \pm    52 $ & $ 1629 \pm   145 $ & $  2.76^{+  0.85}_{ -0.76}$ \\
     &H$\beta$& $  12.5^{+   6.6}_{  -5.5}$ & $ 13^{+  9}_{ -7}$ &
$  948 \pm   113 $ & $ 1375 \pm   357 $ & $  2.20^{+  1.27}_{ -1.10}$ \\
 PG0026+129 &H$\alpha$& $  98.1^{+  28.3}_{ -25.5}$ & $ 95^{+ 39}_{-32}$ &
$ 1961 \pm   135 $ & $ 1117 \pm   109 $ & $73.7^{+23.6}_{-21.7}$ \\
  &H$\beta$ & $ 111.0^{+  24.1}_{ -28.3}$ & $ 92^{+ 32}_{-19}$ &
$ 1773 \pm   285 $ & $ 1719 \pm   495 $ & $ 68.2^{+26.4}_{ -28.0}$ \\
 PG0052+251 &H$\alpha$& $ 163.7^{+  58.5}_{ -38.3}$ & $168^{+ 82}_{-77}$ &
$ 1913 \pm    85 $ & $ :2682 \pm   453 $ & $117.0^{+ 43.1}_{-29.3}$ \\
  &H$\beta$& $  89.8^{+  24.5}_{ -24.1}$ & $ 68^{+ 16}_{-12}$ &
$ 1783 \pm    86 $ & $ 4165 \pm   381 $ & $ 55.7^{+ 16.1}_{-15.9}$ \\
  &H$\gamma$& $  81.6^{+  19.5}_{ -15.9}$ & $ 65^{+ 10}_{ -6}$ &
$ 2230 \pm   502 $ & $ 5633 \pm  3585 $ & $ 79.2^{+ 40.3}_{-38.8}$ \\
 Fairall 9  &H$\beta$& $  17.4^{+   3.2}_{  -4.3}$ & $ 18^{+  3}_{ -4}$ &
$ 3787 \pm   197 $ & $ 6901 \pm   707 $ & $ 48.7^{+ 10.3}_{-13.0}$ \\
\enddata
\tablenotetext{1}{Measurements for blended line.}
\tablenotetext{2}{Measurements based on unblended half of line.}
\end{deluxetable}

\tabletypesize{\scriptsize}
%
%
\begin{deluxetable}{llcccccc}
\tablewidth{0pt}
\tablecaption{Virial Relationship Fits}
\tablehead{
\colhead{} &
\colhead{} &
\multicolumn{3}{c}{Free Slope} &
\multicolumn{1}{c}{ } &
\multicolumn{2}{c}{Fixed Slope $b=-0.5$} \\
\multicolumn{2}{c}{Data Set} &
\multicolumn{1}{c}{$a$} &
\multicolumn{1}{c}{$b$} &
\multicolumn{1}{c}{$\chi^2_\nu$}  &
\multicolumn{1}{c}{ } &
\multicolumn{1}{c}{$a$} &
\multicolumn{1}{c}{$\chi^2_\nu$} \\
\cline{3-5}
\cline{7-8} \\
\multicolumn{2}{c}{(1)} &
\colhead{(2)} &
\colhead{(3)} &
\colhead{(4)} &
\multicolumn{1}{c}{ } &
\colhead{(5)} &
\colhead{(6)} 
}
\startdata
\multicolumn{2}{l}{NGC 5548: RMS (Fig.\ 3)} \\
   & $\sigma_{\rm line}$ vs.\ $\tau_{\rm cent}$ 
   & $3.926\pm0.082$ & $-0.497\pm 0.070$ & 4.04 
& \multicolumn{1}{c}{ } 
   & $3.930\pm0.016$ & 3.86 \\

   & FWHM vs.\ $\tau_{\rm cent}$ 
   & $4.262\pm0.133$ & $-0.427\pm 0.112$ & 4.25 
& \multicolumn{1}{c}{ } 
   & $4.346\pm0.021$ & 4.12 \\

   & $\sigma_{\rm line}$ vs.\ $\tau_{\rm peak}$ 
   & $4.033\pm0.138$ & $-0.583\pm0.117$ & 2.96 
& \multicolumn{1}{c}{ }
   & $3.936\pm0.018$ & 2.88 \\

   & FWHM vs.\ $\tau_{\rm peak}$ 
   & $4.187\pm0.102$ & $-0.378\pm 0.086$ & 2.09 
& \multicolumn{1}{c}{ }
   & $4.328\pm0.020$ & 2.13 \\

 \multicolumn{2}{l}{NGC 5548: Mean\tablenotemark{1}\, (Fig.\ 4)} \\
   & $\sigma_{\rm line}$ vs.\ $\tau_{\rm cent}$ 
   & $4.058\pm0.135$ & $-0.545\pm 0.114$ & 7.60 
& \multicolumn{1}{c}{ }
   & $4.005\pm0.017$ & 7.30 \\

   & FWHM vs.\ $\tau_{\rm cent}$ 
   & $6.202\pm1.602$ & $-2.277\pm 1.357$ & 14.42 
& \multicolumn{1}{c}{ }
   & $4.108\pm0.041$ & 29.15 \\
\enddata
\tablenotetext{1}{Includes narrow-line components.}
\tablenotetext{2}{Narrow component of H$\beta$ removed.}
\tablenotetext{3}{Free-slope fit is unweighted.}
\tablenotetext{4}{Full line width.}
\tablenotetext{5}{Unblended wing of line.}
\end{deluxetable}

\tabletypesize{\footnotesize}

%
%
\begin{deluxetable}{lccc}
\tablewidth{0pt}
\tablecaption{Adopted Virial Products and Derived Black-Hole Masses}
\tablehead{
\colhead{} &
\colhead{$\langle c\tau_{\rm cent}\sigma_{\rm line}^2/G\rangle$} &
\colhead{$M_{\rm BH}$\tablenotemark{1}} &
\colhead{$\log \lambda L_{\lambda}({\rm optical})$}\\
\colhead{Object} &
\colhead{($10^6\,M_{\odot}$)} &
\colhead{($10^6\,M_{\odot}$)} &
\colhead{(ergs s$^{-1}$)}\\
\colhead{(1)} &
\colhead{(2)} &
\colhead{(3)} &
\colhead{(4)} 
}
\startdata
Mrk 335 & $2.58\pm0.67$                 &  $  14.2 \pm 3.7$ 
        & $43.86\pm 0.04$\\
PG 0026$+$129 & $71.4 \pm17.4$          &  $  393 \pm    96$
        & $45.02\pm0.06$\\
PG 0052$+$251 &$67.1 \pm13.8$           &  $  369 \pm    76$
        & $44.96\pm0.08$\\
Fairall 9 & $46.3\pm 10.1$              &  $  255 \pm    56$
        & $44.25\pm0.05$\\
Mrk 590 & $8.64\pm1.34$                 &  $  47.5 \pm  7.4$
        & $43.81\pm0.09$\\
3C 120 & $10.1^{+5.7}_{-4.1}$           &  $55.5^{+31.4}_{-22.5}$ 
        & $44.17\pm0.08$\\      
Akn 120 &$27.2\pm3.5$                   &  $  150 \pm    19$
        & $44.23\pm0.09$\\
Mrk 79 &$9.52\pm2.61$                   &  $    52.4 \pm14.4$
        & $43.72\pm0.05$\\
\enddata
\tablenotetext{1}{Assuming $f=5.5$}
\end{deluxetable}

\end{document}